\documentclass[11pt,a4paper]{article}

\usepackage[utf8]{inputenc}
\usepackage[T1]{fontenc}
\usepackage{textcomp}

\usepackage{amsmath}
\usepackage{amssymb}

\usepackage{graphicx}
\usepackage{float}

\usepackage{booktabs}
\usepackage{array}
\usepackage{multirow}
\usepackage{longtable}
\usepackage{tabularx}

\usepackage[margin=1in]{geometry}
\usepackage{setspace}

\usepackage{microtype}

\usepackage{xcolor}

\usepackage{enumitem}

\usepackage[most]{tcolorbox}

\usepackage[hyphens]{url}

\usepackage[colorlinks=true,linkcolor=blue,citecolor=blue,urlcolor=blue,breaklinks=true]{hyperref}

\title{Europe and the Geopolitics of AGI: The Need for a Preparedness Plan}

\author{
  Maximilian Negele\thanks{Corresponding authors, equal contribution.}$^{1,4}$,\quad
  Daan Juijn\footnotemark[1]$^{2}$,\quad
  Afek Shamir$^{1}$,\quad
  David Jank\r{u}$^{2}$,\\[2pt]
  Beng\"{u}su \"{O}zcan$^{2}$,\quad
  Lisa Soder$^{1,3}$,\quad
  Lucia Velasco$^{8}$,\quad
  Max Reddel$^{2}$,\\[2pt]
  Michiel Bakker$^{5}$,\quad
  Lorenzo Pacchiardi$^{6}$,\quad
  Maksym Andriushchenko$^{7}$
  \\[6pt]
  \small $^{1}$\,RAND Center on AI, Security, and Technology \quad
  $^{2}$\,Centre for Future Generations \\
  \small $^{3}$\,Stanford University \quad
  $^{4}$\,University of Oxford \quad
  $^{5}$\,Massachusetts Institute of Technology \\
  \small $^{6}$\,University of Cambridge \quad
  $^{7}$\,ELLIS Institute T\"{u}bingen and Max Planck Institute for Intelligent Systems \\
  \small $^{8}$\,Oxford Martin AI Governance Initiative
}

\date{December 2025}

\begin{document}
\maketitle
\vspace{-1.5em}

\renewcommand{\thefootnote}{\arabic{footnote}}
\setcounter{footnote}{0}

\begin{center}
\fbox{\parbox{0.92\textwidth}{\small\raggedright
\textit{This paper was originally published as a RAND Europe report (RRA4636-1). \textcopyright\ 2025 RAND Corporation and Centre for Future Generations. This version is redistributed on arxiv with permission under CC BY-NC-ND 4.0. The content is identical to the original apart from formatting and the omission of the executive summary. The original report is available at} \url{https://www.rand.org/t/RRA4636-1}.}}
\end{center}
\vspace{-0.5em}

\begin{abstract}
Artificial general intelligence (AGI)---defined here as AI systems that match or exceed humans at most economically useful cognitive work---has moved from speculation to the centre of political and strategic debate. This paper examines three questions: how soon AGI might emerge, how it could reshape geopolitics, and whether Europe is adequately prepared. Drawing on empirical trends in AI capabilities, expert forecasting surveys, and policy analysis, we find that a plausible window for AGI emergence falls between 2030 and 2040, or potentially earlier, though substantial uncertainty remains. Our analysis of the geopolitical implications suggests that AGI could fundamentally alter the global distribution of economic and military power, intensify interstate competition, and strain existing governance frameworks. Assessing Europe's current positioning, we identify critical gaps: limited strategic awareness of frontier AI progress, structural weaknesses in compute infrastructure and talent retention, low rates of industrial AI adoption, and fragmented policy responses at both EU and Member State levels that do not match the potential scale of disruption. These findings point to a need for a coordinated European preparedness agenda. We outline policy options centred on building institutional capacity for AGI situational awareness, strengthening Europe's position in the AI value chain, and developing frameworks for international stability in an era of increasingly capable AI systems.
\end{abstract}


{
\small
\setlength{\parskip}{0pt}
\vspace{-0.5em}
\tableofcontents
}
\newpage

\section{Introduction}
\label{sec:introduction}

In recent years, artificial general intelligence (AGI) has moved from speculation to the centre of political debate. Senior European policy makers now speak openly about AI systems approaching human-level capabilities on short timescales. In her speech at the Annual EU Budget Conference 2025,\footnote{European Commission (2025e).} Commission President Ursula von der Leyen remarked that when the current budget was negotiated ``we thought AI would only approach human reasoning around 2050. Now we expect this to happen already next year.'' The Commission's AI Continent Action Plan similarly anticipates that ``the next generation of frontier AI models'' could amount to a qualitative leap ``towards Artificial General Intelligence (AGI) capable of tackling highly complex and diverse tasks, matching human capabilities''.\footnote{European Commission (2025b).} Across governments, industry and the expert community, the assumption that AGI may arrive soon is increasingly taking hold.

Another claim has gained traction: that AGI would not merely be another technology but a force capable of reshaping geopolitics.\footnote{Burdette \& Demelash (2025).} Experts increasingly discuss the possibility that more capable AI systems could alter growth trajectories,\footnote{Potlogea \& Ho (2025).} change the balance of military power,\footnote{Burdette \& Demelash (2025b).} and introduce new levers for economic coercion, surveillance and influence.\footnote{Farrell \& Newman (2019).} Indeed, export controls on advanced semiconductors,\footnote{Heim (2025).} the emergence of AI-enabled military systems on the battlefield in Ukraine,\footnote{Bondar (2025).} and large-scale public and private investments\footnote{Soni et al.\ (2025); Abyazov (2025).} in computing infrastructure are now widely interpreted as signs that AI is becoming a central domain of strategic competition.

Experts disagree on the likelihood of short AGI timelines, as well as the extent to which AGI would alter existing power structures. Indeed, some have argued that the rapid growth in AI investment is mostly a reflection of industry hype and amounts to the creation of a new financial bubble.\footnote{Marcus (2025b).} Others have tried to diffuse this argument by pointing at real progress in benchmarks that measure AI capabilities and the rapid revenue growth reported by leading AI companies.\footnote{Wildeford (2025a).} A third camp has pointed out that relatively short timelines to AGI systems are not incompatible with the bubble bursting---much like, after the dotcom bubble burst, the internet did still fundamentally alter society at large.\footnote{Hashim \& Ford (2025).} More fundamentally, the very term of `AGI' is contested, with some shying away from using it,\footnote{Bengio et al.\ (2025).} or preferring alternatives, such as `transformative AI'.\footnote{Agrawal, Korinek, \& Brynjolfsson (2025).}

Although it is too early to know which sides of these debates will prove correct, it is evident that if anything resembling AGI were to emerge soon, Europe would enter this transition under considerable strain. Russia's illegal invasion of Ukraine has reintroduced large-scale land war to the continent and stretched European defence and industrial capacity. Economic stagnation and political fragmentation complicate long-term investment and reform. Transatlantic relations have frayed,\footnote{Munich Security Conference (2025); Vance (2025).} with recent episodes in trade,\footnote{Bown (2025).} security\footnote{Tankersley et al.\ (2025).} and technology policy\footnote{Heim (2025).} underscoring Europe's dependence on US security guarantees, energy supplies and digital infrastructure. Meanwhile, China's material support for Russia and its weaponisation of economic dependencies illustrate how technological and industrial leverage can be used for strategic ends.\footnote{Kirichenko (2025); Baskaran \& Schwartz (2025).} In a world where military capability, economic coercion and technological dominance increasingly determine the international hierarchy, Europe finds itself structurally disadvantaged, possessing neither the military might and innovative edge of the US nor the industrial scale, talent pool and centralised decision-making capacity of China.\footnote{Draghi (2024); Wang \& Kroeber (2025).}

Europe cannot afford to approach the prospect of AGI as a mere regulatory challenge. It must assess whether its current economic, technological and institutional foundations are fit for a world in which AI systems might match or exceed human capabilities across most domains. Technological dependence on foreign AI systems, destabilising arms race dynamics,\footnote{Burdette \& Demelash (2025a).} and the difficulty of ensuring lawful, ethical and robust AI development could leave Europe more vulnerable than at any time in the post-war period.\footnote{Jank\r{u} et al.\ (2024).} Conversely, securing access to enough compute, data, talent and risk-tolerant capital could help unlock new growth, strengthen defence, and allow Europe to shape global norms for safe and ethical AI.

This paper first scrutinises the underlying assumptions behind prevalent statements that AGI is coming soon and that its emergence would reshape geopolitics. We do not aim to provide a comprehensive or definitive treatment of AGI timelines or geopolitical impacts; instead, we conduct a focused review of the most influential arguments and evidence bearing on these two claims, clarifying what is known, what is contested, and where uncertainty remains. We also discuss how different communities define and dispute the term `AGI'. Taking seriously the possibility that AGI could emerge on a relatively short timescale, the paper goes on to assess Europe's preparedness along three key dimensions: strategic awareness, competitive positioning and leverage, and robustness of current policy strategies.

We conclude that Europe is currently underprepared for a transformation to a world with AGI. On present trajectories, the EU and its Member States risk being forced into a choice between deep dependence on foreign AI systems and relative economic and strategic marginalisation. Avoiding this dilemma will require a preparedness plan that addresses three core questions: how can Europe capture economic benefits from AGI while remaining sovereign, how can it prepare its institutions and societies for rapid change resulting from AGI, and how can it strengthen international stability, security and shared prosperity in a world shaped by AGI.

\subsection{Methodology}

\subsubsection{Data collection}

This report is based on a narrative literature review drawing on approximately 300 materials, including peer-reviewed articles, working papers and preprints from research institutions, reports from governments and international organisations, and industry analyses. Relevant material was identified through keyword-based searches in academic databases and search engines such as Google Scholar, and the use of existing bibliographies. The review was conducted iteratively between July and November 2025, with search and synthesis repeated until additional sources yielded diminishing marginal returns. We focused primarily on work from the past 5 years to capture current thinking and recent developments, while incorporating older contributions where they remain analytically relevant. Publications appearing after November 2025 are not included in the research.

We prioritised sources that: (1) directly address at least one of the report's guiding questions on AGI timelines, geopolitical impacts, or Europe's preparedness; (2) offer analytical rather than purely speculative contributions, for example empirically rigorous analyses of frontier capabilities and inputs, or theoretically grounded studies of economic or military implications; (3) meet basic credibility thresholds in terms of venue, author expertise, or transparency of methods and assumptions; and (4) add distinctive evidence or perspectives, including from non-European actors relevant to Europe's choices. The core review is restricted to English-language publications.

\subsubsection{Analytical framework}

Chapter~\ref{sec:how-far-away} combines two complementary analytical lenses on AGI timelines. First, existing forecasts and models---expert surveys, forecasting-platform aggregates, and quantitative models that make explicit assumptions about input factors---are synthesised to obtain an `outside view' on when AGI-level systems might arrive. Second, a plausibility check is conducted based on constraints in the `AI triad' of compute, data and algorithms. Drawing from empirical trends in each component of the triad, we ask whether there are clear structural bottlenecks that would make continued progress towards AGI-level capabilities implausible on a short timescale.

Chapter~\ref{sec:geopolitics} analyses academic, policy and technical work that makes claims about the geopolitical consequences of AGI and classifies these claims into two recurring themes: shifts in economic growth and military capabilities and shifts of power to individuals or AI systems. The chapter summarises arguments in favour of and against each type of shift, highlights major disagreements, filters for mechanisms that appear especially relevant from a European perspective, and documents the reactions of major AI powers to the perceived implications of AGI.

Chapter~\ref{sec:europe-prepared} assesses European AGI preparedness using a three-pronged analytical framework. Drawing from literature in strategic foresight, societal transitions and geopolitics, preparedness is defined as a combination of: (1) strategic awareness in government, meaning the ability of European and national institutions to monitor frontier AI developments; (2) competitive positioning and leverage, meaning Europe's structural position in the global AI value chain; and (3) the robustness of current policy strategies, including their coherence, resourcing and institutional integration.

\section{How far away is artificial general intelligence?}
\label{sec:how-far-away}

Timelines matter in any geopolitical assessment of AGI. If approximately human-level systems can plausibly emerge within the next two decades, investment, institutional reform and international strategy must reflect that possibility. If such systems only emerge much later, or never, the policy response would differ.

The first part of this chapter assembles the main strands of evidence relevant to AGI timelines, using a working definition of AGI as `artificial intelligence (AI) systems matching or exceeding humans at most economically useful cognitive work'. We begin by describing what today's frontier AI systems can already do and where their capabilities remain limited.\footnote{For readers seeking more detail, we recommend consulting Chapter 1 of the International AI Safety Report (Bengio 2025), which gathers the collective insights of experts nominated by the governments of 30 countries, the UN, the EU and the Organisation for Economic Co-operation and Development (OECD).} We then review the existing forecasting and modelling literature on AGI timelines and examine recent progress and potential bottlenecks through the lens of the `AI triad' of compute, data and algorithms.

We find the term AGI useful, which is why we use it throughout this report. We also recognise that it is contested, used inconsistently, and sometimes outright rejected. The second part of this chapter therefore examines these competing definitions and milestones to clarify what various authors mean when they refer to AGI.

This review does not aim to provide a precise prediction. Instead, it assesses whether the emergence of AGI in the relatively near future is sufficiently plausible and consequential that Europe should treat it as a serious planning scenario. In doing so it lays the foundation for the geopolitical analysis and preparedness assessment in the remainder of the report.

\subsection{What today's frontier AI systems can do}

In 2020, early large language models (LLMs) such as OpenAI's GPT-3 impressed observers with fluent text generation. However, their limitations were substantial. They could not properly follow human instructions and processed only text---not images, audio or video. Their context windows were limited, preventing users from inputting large text files, and they were slow and expensive to operate.\footnote{Brown et al.\ (2020).} Without the ability to `think aloud', browse the web or use other tools such as code interpreters, they could not autonomously perform multi-step tasks. Perhaps most significantly, their raw `intelligence' was deficient: GPT-3 struggled with tasks as simple as creating rhymes or multiplying three-digit numbers.\footnote{AI Digest (2025b).}

\begin{figure}[htbp]
  \centering
  \includegraphics[width=0.85\textwidth]{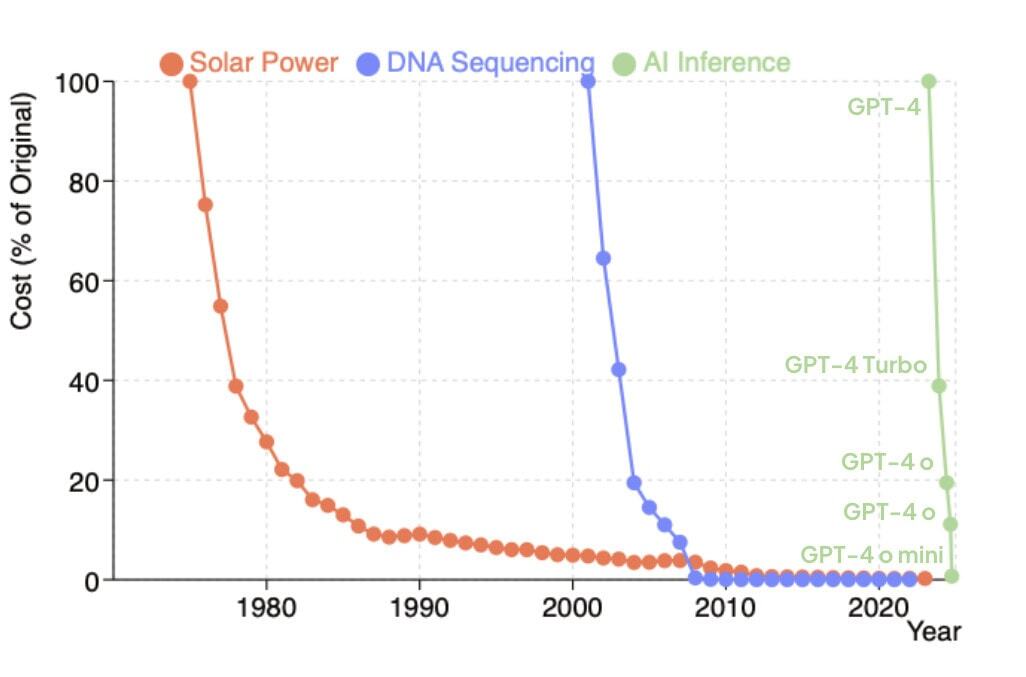}
  \caption{LLM inference price development across tasks. Source: Cottier (2025).}
  \label{fig:inference-prices}
\end{figure}

Five years later, the AI industry has achieved both significant incremental progress and qualitative leaps. Today's frontier systems are multimodal: they process text, audio, images and video, with some generating multiple output types through unified models, including photorealistic images and fluent speech.\footnote{Examples include Gemini 2.5 Flash Image and GPT-4o.} They employ various tools to search the web, use code interpreters, and access external APIs. They can produce chains-of-thought for extended problem-solving and maintain records of previous actions to enable complex, longer-duration software projects.\footnote{Kwa et al.\ (2025).} These qualitative improvements come with advances in speed, affordability (see Figure~\ref{fig:inference-prices}) and raw `intelligence'. Leveraging these capabilities, current models achieve gold-medal performance at the International Mathematics Olympiad,\footnote{Metz (2025).} rank among the world's top competitive coders, achieve superhuman scores on PhD-level science tests,\footnote{Epoch AI (2025b).} and exceed the performance of human doctors on certain diagnostic benchmarks.\footnote{Kolata (2024).} Specialised agentic models can autonomously restructure entire codebases or create functional front-end applications from single prompts, whilst desk research agents can generate fully sourced analytical reports within ten minutes.\footnote{Examples include Claude Code, GPT-5, and Gemini Deep Research, respectively.}

However, while AI systems excel at tests that are challenging for most humans, they remain brittle in many ways. Models frequently hallucinate facts,\footnote{Metz \& Weise (2025).} struggle with simple visual reasoning tasks such as recognising a fourth stripe added to an Adidas logo,\footnote{Wang et al.\ (2024).} misunderstand questions involving basic real-world physics and social intelligence, and fall for logical trick questions.\footnote{Williams \& Huckle (2024).} Models also consistently underperform in areas requiring tacit knowledge or long-horizon planning and execution.\footnote{As noted in Bengio (2025), 171--2.} For example, while Anthropic's Claude can sometimes automate hours of work challenging for junior software engineers, it cannot reliably run a small vending machine business.\footnote{As attempted by Anthropic's Project Vend; Anthropic (2025b).} While Gemini 2.5 Pro excels at difficult mathematics examinations, it fails in peculiar ways when given open-ended goals such as selling online t-shirts.\footnote{AI Digest (2025a).} Meanwhile, progress on physical tasks has lagged behind cognitive capabilities. Although robotic hardware continues to improve, the autonomous operation of robotic bodies remains challenging for AI systems; impressive demonstrations often rely on tele-operation.\footnote{Examples include Tesla's Optimus hand demo.}

\begin{figure}[htbp]
  \centering
  \includegraphics[width=0.85\textwidth]{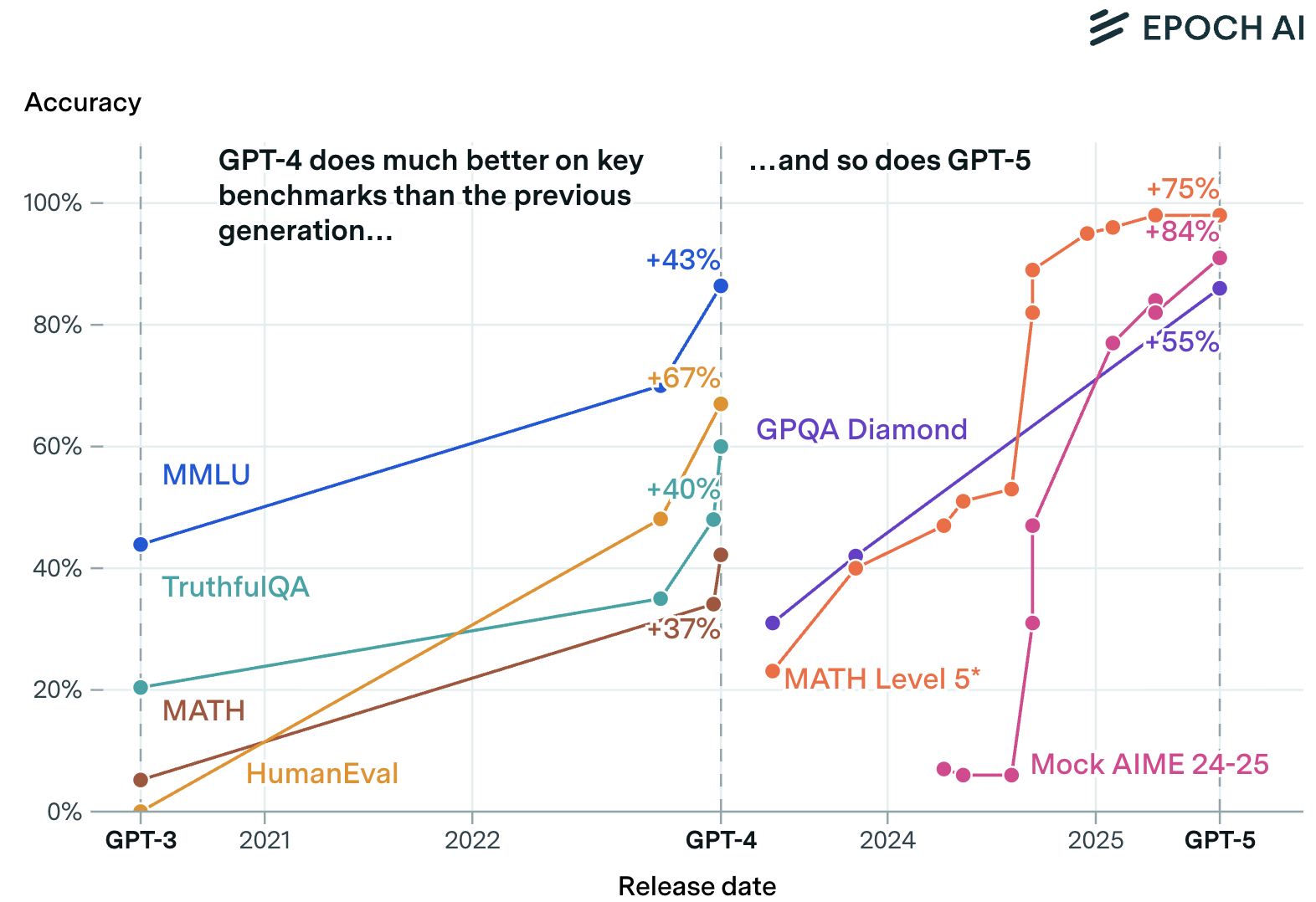}
  \caption{Benchmark performances of GPT3, GPT-4 and GPT-5 compared. Source: Emberson (2025).}
  \label{fig:gpt-benchmarks}
\end{figure}

These contrasting results create what researchers term a `jagged frontier' of AI capabilities, a landscape where models excel in domains with clean, abundant, machine-readable data (such as coding and formal mathematics) but often struggle with experiential knowledge, common sense, or ambiguous, unstructured problems.\footnote{Dell'Acqua et al.\ (2023).} This jagged frontier, combined with industry hype, has made it difficult for occasional observers to assess the capabilities of current AI systems and anticipate future developments. This became particularly evident following the release of GPT-5---a series of models that many commentators found underwhelming, leading some to declare that AI is hitting a wall.\footnote{Marcus (2025a).} However, GPT-5 may have primarily been a product release aimed at raising the capability floor (the quality of tools free users have access to) rather than the ceiling (the maximum capabilities for power users). It scores on-trend on several widely used capability benchmarks\footnote{Wildeford (2025a); see METR (2025b) for evidence of GPT-5's performance on key benchmarks.} and recent research further suggests that the capability jump from GPT-3 to GPT-4 was comparable to that from GPT-4 to GPT-5, but that the latter has been obscured by numerous intermediate releases (see Figure~\ref{fig:gpt-benchmarks}).\footnote{Emberson (2025).}

This suggests that progress is best assessed through longer-term trends rather than individual releases. However, just as capabilities should not be judged solely in the vacuum of benchmarks, they also cannot be inferred directly from real-world adoption rates---a necessary nuance given that adoption data is rapidly evolving. For instance, early results on agentic AI in white-collar workflows remain inconclusive: while some surveys hint at productivity boosts,\footnote{IBM (2025).} others report negligible impact, with some citing unstructured corporate data as the primary bottleneck.\footnote{Estrada (2025); Hill (2025).} Moreover, assessments of progress are further complicated by copyright lawsuits and child safety concerns,\footnote{Heaven (2025); McCallum (2025).} where compliance becomes a prerequisite for products to remain in the market or exist in the first place. While these frictions complicate market adoption and return cycles for AI investments, they do not necessarily indicate a technological plateau.

\subsection{Will rapid progress in frontier AI scale towards AGI?}

If today's systems already perform impressively on many cognitively demanding tasks, the next question is whether this trajectory can lead to capabilities that meet a reasonable definition of AGI. Recent years have seen rapid, sometimes surprising jumps in capability, driven by larger training runs, better data and algorithmic improvements. But it is not obvious whether this pace can be maintained, or whether economic, technical or physical constraints will force a sharp slowdown before systems reach AGI-level performance.

In this section, we examine the prospects for further progress. We first review the main forecasting and modelling efforts that translate current trends and expert judgement into explicit timelines for AGI or `transformative AI'. We then make an assessment of the underlying drivers of progress, using the `AI triad' of compute, data and algorithms. The aim is to evaluate whether the literature suggests clear structural reasons---such as binding limits on energy, data or algorithmic improvement---that would make the continued advance towards AGI implausible.

For the purpose of this chapter we define AGI as `AI systems that can match or exceed humans across most economically relevant cognitive tasks'. The final section of this chapter discusses different definitions of AGI in more detail and outlines the different milestones associated with each.

\subsubsection{Assessing evidence from expert and model-based AGI forecasts}

A useful starting point for an outside-view assessment of AGI timelines is the recent synthesis of expert forecasts by Todd (2025), which reviews five different forecasting communities: leaders of frontier AI companies, surveyed AI researchers, the Metaculus prediction platform, a 2022 survey of `superforecasters' (the XPT survey), and the Samotsvety forecasting group.\footnote{Samotsvety is a group of expert forecasters, founded by Misha Yagudin, Nu\~{n}o Sempere and Eli Lifland.} Taken together, these sources show a marked shift towards shorter timelines over the last few years, and place the emergence of AGI before 2030 clearly within the range of serious expert opinion, even though many experts still predict that AGI will emerge later.

Leaders of major AI companies have increasingly suggested that AGI could arrive within roughly 2--5 years.\footnote{Tech Desk (2024); Altman (2025); CNBC Television (2025); Hassabis (2024); Kantrowitz (2025).} These estimates are subject to significant selection and incentive effects, but the same individuals have unusually good visibility into upcoming model generations and were comparatively accurate in anticipating past capability jumps. Todd argues that their views are likely optimistic, but perhaps only by several years rather than decades. Surveys of a much broader set of AI researchers indicate a somewhat longer timeline: in a 2023 survey of AI researchers, the median respondent assigned a roughly 25 per cent probability to `high-level machine intelligence' by the early 2030s and 50 per cent by the 2040s, while also sharply revising many task-level forecasts forward after seeing the performance of large language models.\footnote{Grace et al.\ (2024).}

At the same time, the researchers in this group have been repeatedly surprised by near-term progress and give internally inconsistent answers, which suggests that one should treat their quantitative estimates with caution rather than as precise guidance.\footnote{Grace et al.\ (2024).} Professional forecasters and forecasting platforms show a similar pattern of shortening timelines. On Metaculus, a long-running community forecast using a relatively demanding AGI definition (including general robotics), the timeline for AGI moved from a median of about 50 years away in 2020 to a roughly 25 per cent chance by 2027 and 50 per cent by 2031, as of late 2024.\footnote{Metaculus (2025).}

The 2022 XPT survey of superforecasters, which predates some of the latest progress, produced substantially later medians,\footnote{Karger et al.\ (2023).} but more recent judgement-based work by the Samotsvety group points to much shorter timelines again, with estimates of a roughly 25 per cent chance of AGI by the late 2020s.\footnote{Wynroe (2023).} Todd's overall conclusion is that expert opinion has moved noticeably towards shorter timelines; the various forecasts neither rule out nor confirm near-term AGI, but jointly support treating a 3--15 year horizon as a live possibility.

Complementing this evidence, a recent article by Wynroe (2023) for Epoch AI provides a structured literature review of transformative AI (TAI) timelines, focusing on explicit quantitative models and prominent judgement-based forecasts. While TAI is not identical to our notion of AGI, it is defined similarly as AI with impacts on the order of the industrial revolution, and the underlying forecasts are highly relevant to our question. Epoch distinguishes `model-based' forecasts---which specify an explicit arrival-time model---and `judgement-based' forecasts from individuals or forecasting communities. They then elicit internal weights over these forecasts and aggregate them to obtain summary distributions over TAI/AGI arrival dates.

Among model-based forecasts, inside-view approaches such as Ajeya Cotra's bioanchors model\footnote{An attempt to forecast the development of transformative AI anchored in estimates of computational processes found in evolution and the human brain.} assign substantial probability mass this century,\footnote{Cotra (2020).} with a median AGI arrival date around the 2050s (e.g.\ ${\sim}8$ per cent by 2030, ${\sim}47$ per cent by 2050, ${\sim}78$ per cent by 2100).\footnote{Epoch AI (2025d).} Outside-view models such as `semi-informative priors'\footnote{An attempt to forecast the development of AGI based on a series of `success-failure' stages, informed by AI input factors.} and `insight-based' approaches typically yield longer timelines, with medians in the second half of the century or later.\footnote{Epoch AI (2025e).} When Epoch aggregates the model-based work using their subjective weights, they obtain roughly an 8 per cent probability of TAI/AGI by 2030, ${\sim}27$ per cent by 2050, and ${\sim}54$ per cent by 2100, with a median around 2080--2090.\footnote{Wynroe (2023).}

The judgement-based forecasts reviewed by Epoch tend to be more aggressive than the model-based examples. Aggregating across sources such as the AI Impacts 2022 survey, Metaculus, Samotsvety, and individual forecasters including Cotra and Karnofsky, Epoch's weighted judgement-based summary assigns a roughly 25 per cent chance of TAI/AGI by 2030, ${\sim}57$ per cent by 2050, and ${\sim}88$ per cent by 2100, with a median around the mid-2040s.\footnote{Wynroe (2023).} These figures are broadly consistent with the expert-forecast synthesis by Todd, but make explicit the divergence between more conservative model-based priors and more bullish all-things-considered judgements. It should be noted that the Epoch review was completed in 2023 and therefore does not reflect subsequent advances in frontier AI systems or any post-2023 shifts in expert judgment.

\subsubsection{Assessing plausible constraints on the path to AGI}

\begin{figure}[htbp]
  \centering
  \includegraphics[width=0.7\textwidth]{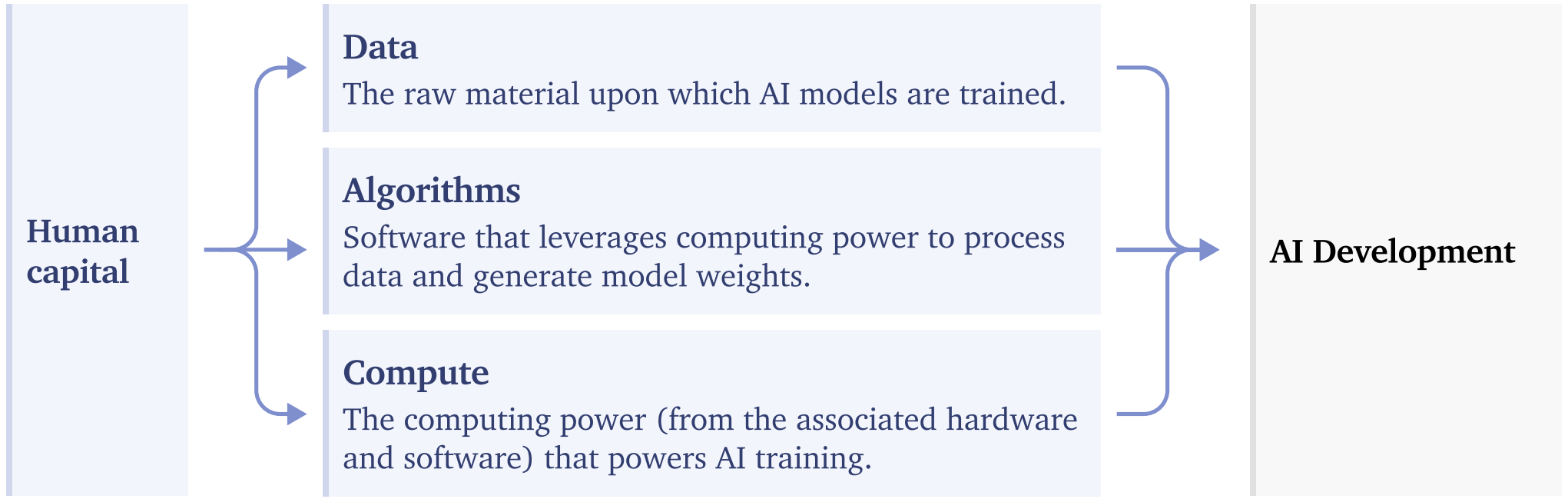}
  \caption{The AI triad. Source: Heim et al.\ (2024).}
  \label{fig:ai-triad}
\end{figure}

The existing forecasting and modelling literature on AGI arrival translates a complex and rapidly evolving technological trajectory into probability distributions. To complement this work, in this section we examine recent AI progress through the lens of the `AI triad' (see Figure~\ref{fig:ai-triad}) as an organising framework for asking whether there are structural reasons to expect progress to stall before AGI-level capabilities are reached.

Rather than reproducing the detailed quantitative structure of existing models, such as those mentioned in the previous section, we ask a simpler question: given current capabilities and observed trends in compute, data and algorithms, does the literature identify any clear, robust bottlenecks that would make near-term AGI implausible? In this sense, the AI-triad analysis serves as a robustness check on the forecasting literature rather than a competing forecast.

Broadly, recent improvements in AI capabilities have resulted from a combination of three input factors that researchers collectively term the `AI triad': data, compute, and algorithms.\footnote{Buchanan (2020).} So-called scaling laws (empirical relationships that have held over many orders of magnitude) demonstrate how increases in two of these inputs---compute and data---reliably lead to increased performance. Recent developments in the triad can be summarised as follows:

\begin{itemize}
  \item \textbf{Data:} The volume of data used to train frontier models has expanded by 3.6 times per year since 2020.\footnote{Epoch AI (2025b).} Data quality and diversity have also improved: models now train on carefully curated datasets spanning books, scientific papers, code repositories, videos and images, providing exposure to broader domains of human knowledge.\footnote{As shown in Google DeepMind's paper on Gemini 2.5; Google Team (2025).} These increases in data quantity, quality and diversity have enabled researchers to productively scale the compute used for AI training.

  \item \textbf{Compute:} The computational power devoted to training frontier models has grown by approximately 5 times annually over the same period.\footnote{Sevilla (2024).} This explosive growth stems from larger AI datacentres, increasingly sophisticated AI chips, and longer training runs.

  \item \textbf{Algorithms:} Advances in algorithmic efficiency (including new training methods and model architectures) mean AI systems can achieve equivalent performance with roughly half the compute every eight months.\footnote{Ho (2024).} Occasionally, breakthrough innovations deliver gains far exceeding these average improvements, for example the shift to the transformer architecture enabling the GPT series, or the recent development of reasoning models.\footnote{Ho (2025).}
\end{itemize}

These inputs are growing at a fast rate: `effective compute'---a combined term for both compute and algorithmic efficiency---increases around 15-fold every year.\footnote{Juijn (2024a).} These combined increases have enabled recent progress in AI capabilities. If the same trend continues, we expect that the jagged capabilities frontier will further expand outwards, potentially surpassing human capabilities across most domains. The remainder of this section assesses the potential for such continued improvements in the three components of the AI triad.

\paragraph{Data: how reinforcement learning may break the data wall.}

For scaling laws to hold, increases in compute must be matched by increases in fresh data.\footnote{Hoffmann et al.\ (2022).} However, by late 2024, AI systems were already training on most of the available high-quality internet data.\footnote{Dubey et al.\ (2024).} This caused some experts to expect that AI progress might soon decelerate significantly due to data scarcity.\footnote{Patel (2023).} This argument, often termed the `data wall', may have proved partially correct: it remains plausible that data scarcity has led to diminishing real-world returns of pre-training---the initial phase of AI training where models mostly learn to imitate the training data.\footnote{Juijn (2024a).} However, since 2024, AI developers have discovered a promising approach to scale beyond the data wall that does not solely rely on scaling pre-training.

\begin{figure}[htbp]
  \centering
  \includegraphics[width=0.85\textwidth]{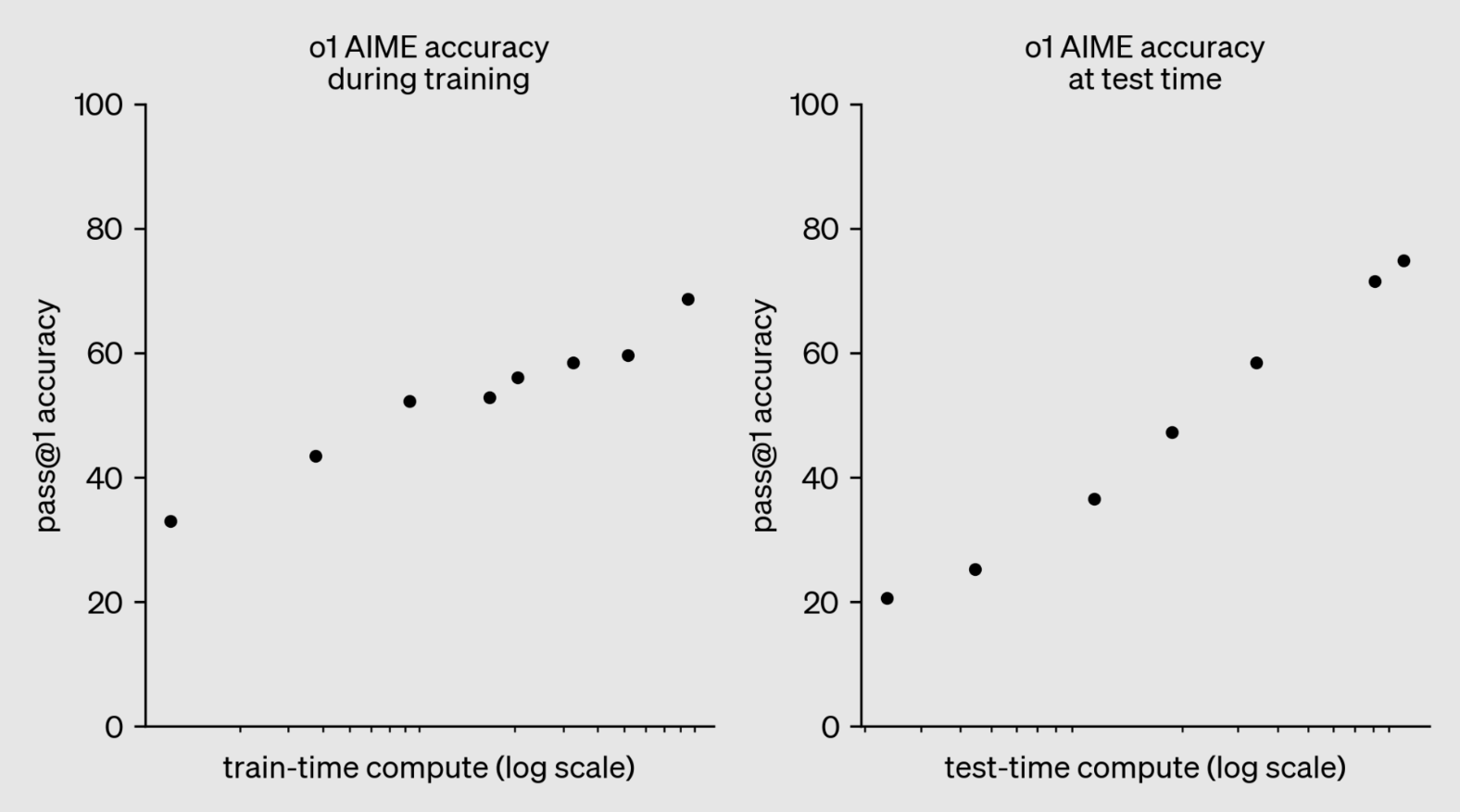}
  \caption{Scaling laws compared across train-time and test-time compute. Source: OpenAI (2024).}
  \label{fig:scaling-laws}
\end{figure}

The recent emergence of reasoning models and agentic AI systems has increased the compute used during the reinforcement learning stage of training LLMs, often termed post-training.\footnote{Examples include OpenAI's o1 and Claude Code.} During this stage, AI models attempt to solve challenging problems by articulating their internal reasoning steps through chains-of-thought---like a scratchpad that allows AI models to `think out loud'. Novel inference-time scaling relationships (distinct from pre-training scaling laws) suggest that the more elaborate these chains-of-thought---and thus the more compute is spent---the better an AI system's answers become, up to a point where returns tail off (see Figure~\ref{fig:scaling-laws}).\footnote{Heim (2024); Ord (2025).} This matters because inference compute can scale independently of training compute and can be aimed specifically at important problems.\footnote{Luong and Lockhart (2025).}\textsuperscript{,}\footnote{For instance, this approach enabled recent models by Google DeepMind and OpenAI to win gold medals in the International Mathematical Olympiad.}

In contrast to training-time scaling, inference-time scaling holds the model fixed and instead allocates more compute per query, for example by running multiple agents in parallel for hundreds of thousands of times.\footnote{Patel and Kourabi (2025).} Each attempt is then evaluated using separate `verifiers'---software programmes that check whether code functions as intended, or other AI models that rate outputs based on carefully constructed rubrics. Many of the `reasoning traces' generated in this process lead to incorrect or suboptimal answers. By learning from the contrast between correct and incorrect reasoning traces, the model gradually recognises promising reasoning directions early, abandons unproductive paths, and develops more robust problem-solving strategies.\footnote{Guo et al.\ (2025).} This deliberate exploration of both successful and failed reasoning paths enables developers to invest substantial compute in generating rich learning signals. Through this process, the model's own reasoning traces---both successful and failed attempts---become training data for future iterations.

This new stage of learning may soften the bottleneck of human data availability. As long as external verifiers can differentiate between better and worse answers, and models can productively use additional compute to `think longer', outputs can be further improved. The advancement of post-training has thus reduced the probability that AI progress will stall due to data unavailability. However, challenges remain. Assuming poor generalisation to tasks outside those tested during training, data constraints may still prove limiting: an AI assistant trained to help Americans file their taxes may struggle to apply those skills in other (e.g.\ European) jurisdictions. This could lead to more measured progress. Furthermore, there may be limits to how much the relative size of post-training can be scaled compared to pre-training before models become over-optimised for easily verifiable problem-solving capabilities.\footnote{Lambert (2025).}

Continual learning could offer another path to ease data bottlenecks, by turning an agent's ongoing experience into a training signal, rather than relying only on finite, human-curated datasets. One research agenda envisions long-lived agents that learn and plan continuously in non-stationary environments, with learning grounded in the stream of observation, action, and reward instead of special training periods or heavy human labelling. If this approach scales, real-world use could supply more task-relevant data and help agents adapt across domains and jurisdictions, though major challenges in robust, safe continual learning remain.\footnote{Sutton et al.\ (2022).}\textsuperscript{,}\footnote{Wang et al.\ (2025).}

In a similar vein, advances in `world models' are removing limits to training data from the real world. Recent examples\footnote{Parker-Holder and Fruchter (2025).} include Google DeepMind's Genie 3 and NVIDIA's Cosmos, which create interactive simulated environments for training and experimentation. Genie 3 can generate navigable worlds from text prompts in real time, while Cosmos provides models and tooling to build customised simulations for robots, autonomous vehicles and other physical AI systems. If these approaches scale, they could supply abundant, task-relevant synthetic experience, reducing the cost and delay of real-world data collection and improving the path from training in simulation to partial transfer on real hardware, currently one of the weakest points in the jagged frontier of AI capabilities.

Taken together, the current evidence does not support a simple `data wall' situation in which progress necessarily stalls for lack of additional human-generated text or images. Over the next few model generations, there still appears to be headroom in terms of high-quality web, code and multimodal data, and synthetic data and AI-generated feedback offer additional ways of extracting more learning signals from existing sources. At the same time, it is unclear how far these techniques can be pushed without running into quality degradation, mode collapse, or blind spots in domains where real-world data remains scarce. Overall, the data situation does not provide a robust argument against continued progress towards AGI, but it does suggest that future gains may depend increasingly on clever use and curation of data rather than sheer volume alone.

\paragraph{Compute: rapid growth despite potential slowdowns.}

Even if training AI systems on ever more data remains possible, securing the required computing resources to do so remains a challenge. With training compute increasing by five times annually,\footnote{Sevilla (2024).} it becomes more and more difficult for leading AI companies to purchase enough AI chips, build datacentres to house those chips, and secure enough power to operate them.

A leading projection estimates that compute scaling could theoretically continue at roughly its current pace through to 2030, given sustained investment and sufficient commercial and political determination.\footnote{Sevilla et al.\ (2024).} This research identified power availability as the greatest potential bottleneck---more significant than chip production capacity. A recent compute forecast reached similar conclusions: given strong demand, the chip supply chain can likely accommodate the continuation of current compute trends.\footnote{Kokotajlo et al.\ (2025).} In this section we examine two of the main potential obstacles to compute scaling: rapidly increasing power requirements and escalating investment volumes. We focus primarily on the US companies leading the technology frontier and dominating datacentre investments.

Recent research has found that AI datacentres could require 68 gigawatts of peak power capacity globally by 2027, an amount slightly larger than the total power capacity of Poland, assuming continued rapid growth in chip supply.\footnote{Estimates included in research by Pilz, Mahmood \& Heim (2025). Poland's grand system capacity is at 64.3 gigawatts in 2025 as per Entsoe (2025b), the EU electricity transparency hub.} These projections raise questions about grid capacity and whether power equipment supply chains can match the sector's growth.\footnote{In March 2025, SemiAnalysis estimated that AI power demand will increase by approximately 70 gigawatts by 2028, with most demand originating from the US; Patel (2024a).} Leading AI companies are already investing in dedicated power infrastructure, including reviving nuclear plants and dismantling foreign plants to ship components to the US, to secure supply.\footnote{For example, a Three Mile Island nuclear plant will reopen to power Microsoft data centres, and xAI bought an overseas power plant to power a one million AI GPUs data centre in the US.} They are also investing in distributed training set-ups, where models can be trained over multiple geographically separated datacentres.\footnote{Patel, Nishball \& Ontiveros (2024).} Yet some AI companies remain energy-constrained: multiple datacentres scheduled for 2025 have allegedly been delayed due to limited power availability.\footnote{Patel, Ontiveros et al.\ (2025).} The quest for energy to power AI datacentres was used as a justification for a recent Middle East arrangement, in which American companies signed multi-gigawatt contracts with Saudi and Emirati entities to rent AI compute from locations with ample spare power capacity. Indeed, securing additional power capacity for AI has become a policy priority in Washington, with the recently launched AI Action Plan dedicating one of its three pillars to AI and energy infrastructure.\footnote{White House (2025).}

Whether government action, multi-datacentre training and foreign investment will sufficiently alleviate energy bottlenecks remains unclear. Companies are employing creative solutions to circumvent constraints, such as refurbishing old cryptocurrency mining facilities,\footnote{Dave (2024).} temporarily powering datacentres with mobile gas generators,\footnote{As xAI's Colossus has shown; see Mann (2025).} or housing chips in large tent-like structures.\footnote{Meta has been reported to be building AI data centres in tents; see Martindale (2025).} Given the lengths to which companies are going to secure power supply and accelerate buildouts, compute scaling is unlikely to halt due to power shortages in the near future. Moreover, planned or partially constructed datacentre projects should enable the industry to scale compute for approximately two more model generations, as evidenced by recently announced megaprojects such as OpenAI's 5 gigawatt Stargate Project and Meta's 5 gigawatt Hyperion.\footnote{OpenAI (2025v) and Zeff (2025), respectively.}

Whether companies can finance increasingly large clusters is a separate question from energy availability. Frontier AI companies have seen revenues soaring and exceeding expectations throughout 2025,\footnote{For example, OpenAI hit \$12bn in annualised revenue this year, according to Reuters (2025a).} but such rapid growth has also fuelled concerns about a potential `AI bubble',\footnote{Schaul \& De Vynck (2025).} where investment and valuation trends outpace actual realised returns. If jagged capability progression slows broad market adoption, investments could theoretically dry up quickly, as was outlined in a recent AI scenario study.\footnote{\"{O}zcan et al.\ (2025).} These financial uncertainties are amplified by the increasing scale and complexity of compute buildouts, which lengthen project lead times.\footnote{Edelman (2025).} This makes intuitive sense: scaling from 1,000 to 10,000 Graphic Processing Units (GPUs) may only require purchasing new servers (assuming spare datacentre capacity), but scaling from 100,000 to 1 million GPUs may require building new power plants, requiring additional preparation and construction time. Before making new multi-billion or even multi-hundred-billion dollar investments, AI companies and investors could pause to confirm that earlier investments are paying off, resulting in a more iterative investment rhythm than in past cycles. However, in the current hyper-competitive landscape, such an iterative strategy may be deemed too slow by leading AI developers.

In light of increasing infrastructure buildout timelines and the possibility of more measured capital allocation, it seems at least plausible that compute growth will decelerate somewhat in the coming years. Nevertheless, given the current extraordinary levels of annual growth, policy makers should expect major increases in compute before the end of the decade. In a recent interview, leading experts expressed confidence that training runs would reach $10^{29}$ FLOP (floating point operations) before 2030, despite energy and investment constraints.\footnote{Sevilla and Edelman also note that compute scaling has thus far been driven by increases in total training time, which they expect to cease soon. Whilst we accept this general argument, we note that an argument can be made in the opposite direction: fast iteration cycles may incentivise companies to skip new pre-training runs, instead focusing on continuing RL-training for longer on the same model, effectively increasing total training time.} This equates to models roughly 1,000 times larger than today's largest model---a scale-up roughly comparable to the compute increase from GPT-3 to GPT-5.

Finally, the above analysis has only addressed training compute so far, rather than inference-time compute: future AI-driven breakthroughs may stem from models trained with more compute, but also from giving models resources to `think longer'.\footnote{Heim (2025).} Assuming further progress in productively harnessing inference-time compute, substantial potential remains for this second scaling approach. Current models typically struggle to make productive use of more than a few euros' worth of compute at once, whereas some problems---such as finding cures for medical diseases---could arguably warrant expenditures exceeding \texteuro{}1m at a time.\footnote{As suggested by Brown (2025).}

In summary, there are clear medium-term economic and physical constraints to growing compute: building ever larger clusters requires substantial capital expenditure, grid upgrades and reliable access to power, and the historic pace of exponential growth in training compute cannot continue indefinitely. Nonetheless, current plans for datacentre and accelerator buildout, combined with the scope for further hardware efficiency gains, make it plausible that total effective training compute can increase by one or more orders of magnitude beyond today's frontier runs. In the absence of strong policy restrictions or a major reversal in investment, compute constraints alone do not yet appear to constitute a decisive barrier to continued progress towards AGI-level systems, though they could slow the pace or concentrate capabilities in the hands of a small number of actors.

\paragraph{Algorithms: potential for further acceleration.}

Of the three components of the AI triad, future algorithmic progress may be most difficult to extrapolate. On the one hand, algorithmic progress could decelerate as ideas are becoming harder to find, for instance because improvements only show up at very large compute scales.\footnote{Josephson (2025); Bloom et al.\ (2017).} Even as the AI industry grows and more researchers seek novel algorithmic breakthroughs, maintaining the current pace of innovation may prove impossible. Moreover, the current technological paradigm might encounter an unexpected dead end, requiring developers to backtrack and restart from an earlier branch of the technology tree.\footnote{Toner (2025b).} On the other hand, the industry might discover another breakthrough comparable to the transformer architecture, unlocking further efficiency improvements.\footnote{As mentioned earlier, an idea that has recently gained traction is continual learning: AI systems that improve continuously based on personalised user interactions. See Denain (2025a), Wang et al.\ (2025), and Sutton, Bowling, and Pilarski (2022).}

An important question is the extent to which AI systems can accelerate their own R\&D. If AI can reliably contribute to designing, testing and training better AI systems, this will create a positive feedback loop.\footnote{Sett (2024).} Each generation of model will help to build the next, potentially compressing development timelines from years to months or even weeks. The likelihood of such scenarios and the extent of the resulting acceleration remain unclear. A recent theoretical study on this topic estimates a 40 per cent chance that three years of capabilities progress could be compressed to merely four months through AI R\&D automation.\footnote{Davidson \& Houlden (2025).} Similarly, the AI 2027 scenario presents a case for rapid AI feedback loops resulting in explosive capability growth and vastly superhuman intelligence.\footnote{Kokotajlo et al.\ (2025).} Other researchers disagree with this picture. For example, an interview study from 2024 involving eight leading AI researchers identified substantial disagreements on timing and acceleration speeds, but also general agreement that core engineering tasks could theoretically be automated.\footnote{Owen (2024).}

Empirical evidence helps to assess the likelihood and importance of automated AI R\&D. In recent years, AI systems have achieved measurable gains at multiple points in the research and development pipeline, from accelerating exploratory experimentation to handling complex machine learning engineering tasks.\footnote{Yang et al.\ (2023); Chan et al.\ (2024).} AI R\&D is unusually amenable to automation compared with many other forms of cognitive work. Training logs, experiment configurations and evaluation metrics are already machine readable. Models can in principle review all past experiments and code changes, whereas human researchers only ever see a small fraction. Reward signals are relatively clean (did the model improve on a benchmark or not) which makes automated search over architectures, hyperparameters and training schemes easier. With abundant data and clear feedback, AI systems can scale across the full surface of an organisation's experiments in a way individual researchers cannot.\footnote{Owen (2024).}

\begin{figure}[htbp]
  \centering
  \includegraphics[width=0.85\textwidth]{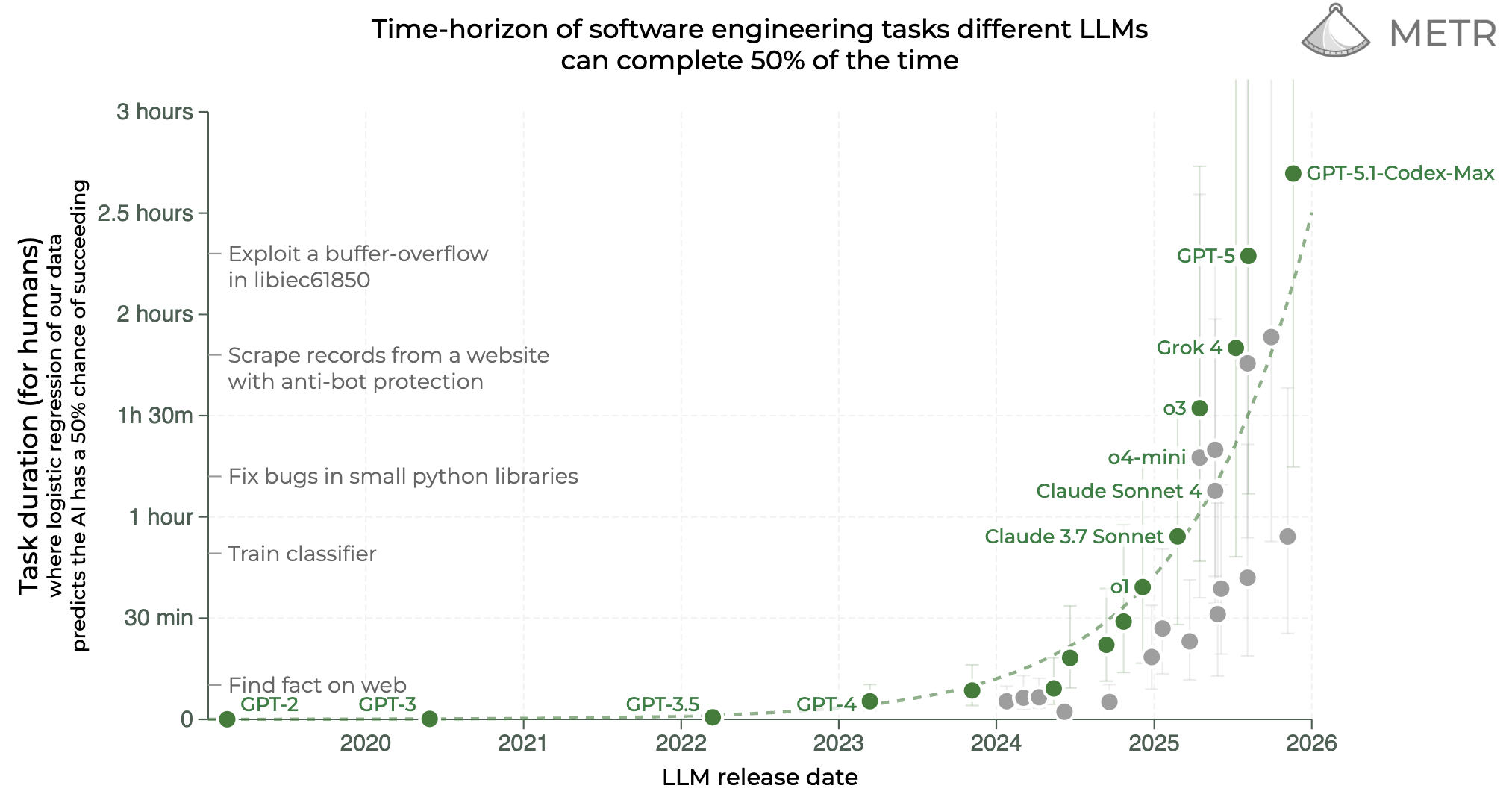}
  \caption{Time-horizon of software engineering tasks different LLMs can complete 50 per cent of the time. Source: Kwa et al.\ (2025).}
  \label{fig:swe-tasks}
\end{figure}

In practice, AI is arguably already generating positive feedback loops by simple coding automation: Anthropic claims that their AI system wrote 80 per cent of the code for their new coding agent, Claude Code.\footnote{Latent Space (2025).} This anecdotal evidence largely aligns with research by AI-evaluation organisation METR,\footnote{Kwa et al.\ (2025). A different METR study (2025c) shows that software engineers may overestimate the productivity benefits that current AI coding tools bring.} which systematically tracked the maximum task complexity of relevant software engineering tasks that AI agents can autonomously perform. This research reveals that every seven months, AI systems can successfully complete tasks twice as difficult, as defined by the time humans require to complete them (see Figure~\ref{fig:swe-tasks}).\footnote{Kwa et al.\ (2025).} More recently, METR has demonstrated that similar relationships hold for other task types including mathematics, self-driving, machine learning research, and answering difficult science questions.\footnote{METR (2025a).} If this trend continues, AI systems would achieve 50 per cent success rates on month-long software engineering tasks by 2031---perhaps sufficient to automate large portions of AI R\&D.

Algorithmic progress is the hardest component of the triad to forecast but has been a powerful driver of capability gains in recent years, sometimes offsetting slower growth in data or compute. Over the past decade, improvements in model architectures, optimisation methods and training schemes have delivered substantial efficiency gains and unlocked qualitatively new behaviours. There is no guarantee that similar breakthroughs will continue, and it is possible that some of the most accessible ideas have already been exploited. At the same time, the growing use of AI systems themselves to assist in research and engineering could cause a further acceleration. Overall, the algorithmic landscape provides no clear reason to expect an imminent, robust slowdown before reaching AGI-level performance; if anything, it adds asymmetry to the picture, with significant upside potential for new techniques to push capabilities further and faster than simple extrapolations imply.

\subsubsection{Conclusion: the near-term emergence of AGI is plausible enough to warrant significant attention from European policy makers}

This section has examined three strands of evidence relevant to AGI timelines. Firstly, we have shown that frontier AI systems already match or exceed human performance on a growing range of cognitively demanding tasks, while remaining brittle in others, resulting in a `jagged frontier' of capabilities. Secondly, we have reviewed expert and model-based forecasts, which, despite wide disagreement and methodological uncertainty, increasingly place a non-trivial probability on AGI or transformative AI in the coming decade, including before 2030. Thirdly, we have assessed potential constraints using the AI triad of compute, data and algorithms, finding significant headroom in each component and no single, robust bottleneck that would make further rapid progress towards AGI implausible.

On balance, we interpret this evidence to suggest that AI systems matching or exceeding humans at most economically valuable tasks could plausibly emerge between 2030 and 2040, and that there is a non-trivial chance of such systems arriving even sooner. We do not claim that AGI is likely or inevitable on any particular timeline, nor that existing models and forecasts are precise enough to justify a narrow confidence interval. Our claim is more modest: \textbf{a near-term AGI scenario is plausible enough to warrant significant attention from European policy makers.}

We emphasise that this assessment comes with significant uncertainty. The forecasting and modelling approaches used and reviewed here have clear limitations. Moreover, there are serious arguments against near-term AGI which we do not dismiss. For example, the International AI Safety Report emphasises conceptual and technical gaps in today's systems and argues that achieving AGI may require multiple paradigm shifts, which might take much longer than historical experience suggests.\footnote{Bengio et al.\ (2025).}

The remainder of this report therefore treats the near-term emergence of AGI as a working scenario. We do not assume that this scenario will materialise, but ask what its geopolitical consequences would be and how prepared Europe would be if it does.

\subsection{What would it mean to achieve AGI?}

AGI definitions are numerous and involve a wide range of qualitative and quantitative milestones. OpenAI defines AGI as `a highly autonomous system that outperforms humans at most economically valuable work', which arguably includes most physical labour.\footnote{OpenAI (2025a).} The International AI Safety Report instead notes that AGI is often described as a system that `equals or surpasses human performance on all or almost all cognitive tasks'---seemingly covering only white-collar work.\footnote{Bengio et al.\ (2025).} Google DeepMind, meanwhile, prefers discussing `levels of AGI', distinguishing between narrow and general capabilities, and between the percentage of humans an AI system outperforms.\footnote{Google DeepMind (2024).} Recent work has tried to create a measurable, empirically grounded framework for AGI. Hendrycks et al.\ (2025) define AGI as `an AI that can match or exceed the cognitive versatility and proficiency of a well-educated adult', operationalising this through the Cattell-Horn-Carroll theory of human cognition. Some discussions also refer to artificial superintelligence (ASI),\footnote{Bostrom (2014), 52--65.} a hypothetical class of systems that would vastly exceed even the most capable humans across all domains, potentially beyond human oversight, and often treated as a speculative upper bound of the AGI spectrum. The term increasingly appears as an explicit goal to be chased by AI developers, such as Meta Superintelligence Labs or Safe Superintelligence Inc.\ (SSI), and in studies exploring AGI trajectories,\footnote{Kokotajlo et al.\ (2025).} but its definitions are even sparser and less concrete than those for AGI in both research and policy contexts.\footnote{Tidler et al.\ (2025).}\textsuperscript{,}\footnote{European Commission (2025m).}\textsuperscript{,}\footnote{Definitions of AGI should be distinguished from definitions of `general-purpose AI models' and `general-purpose AI models with systemic risk'. The latter set a much lower bar, saying that a model `displays significant generality and is capable of competently performing a wide range of distinct tasks'. Hence, a model can be considered GPAI without being capable of economically valuable labour.}

With varied interpretations and no consensus on a preferred threshold, the term AGI creates considerable confusion. Some have claimed that AGI may already exist, whilst others place it decades away.\footnote{Patel (2025a).} Without a shared definition, it is difficult to say where the disagreement comes from. Because the term AGI is so engraved in public discussions about AI progress, it is nevertheless important for policy makers to understand the different flavours in circulation. Below we provide a brief overview of four categories of milestones that are often linked to the achievement of AGI:

\begin{enumerate}
  \item \textbf{Milestones focusing on AI's ability to match or exceed humans at AI research and development (R\&D).} In the AI 2027 scenario, AGI is described as `imminent' when AI systems have become better machine learning researchers than most human experts.\footnote{Kokotajlo et al.\ (2025).} As discussed above, when AI systems can automate most of their own improvement, AI progress could accelerate dramatically. While the AI 2027 authors do not explicitly equate the achievement of the `automated AI researcher' to AGI, the threshold provides an intuitive AGI milestone that is often used by researchers in informal conversations. That said, definitions of AGI that index on AI R\&D do remain relatively narrow and may fail to grasp the intuition that AGI should enable large-scale automation outside of the AI field itself.

  \item \textbf{Milestones focusing on AI's ability to match or exceed humans at non-embodied aspects of broader science and R\&D.} Anthropic CEO Dario Amodei argues that AI's primary contribution to human welfare may be through accelerating science and R\&D, with particular emphasis on biology.\footnote{Amodei (2024) also prefers the term `Powerful AI' over AGI.} When AI systems can match or exceed humans in all non-embodied aspects of such work---reviewing literature, forming novel hypotheses, planning experiments, instructing humans how to physically run them, interpreting outcomes, and revising theories---Amodei argues they could compress years of scientific progress to mere months, functioning as a `country of geniuses in a datacentre'. Counter-intuitively, Nobel Prize-winning AI systems may prove easier to develop than those that can reliably automate most mundane white-collar work, as science could provide an easier training environment. This distinction potentially justifies using the `genius AI scientist' as an appropriate milestone for AGI or `powerful AI'.

  \item \textbf{Milestones focusing on AI's ability to match or exceed humans at cognitive work.} Some researchers have challenged the claim that AI's primary contribution to the economy or human welfare would occur through science and R\&D.\footnote{Erdil (2025b).} They argue instead that economic transformation relies on the broad automation of cognitive work. If correct, a more relevant milestone may be achieved when AI systems can automate all complex cognitive tasks in the broader economy, including those that rely on undocumented tacit knowledge. Such tasks may remain beyond an AI scientist's reach, potentially justifying a more ambitious definition encompassing all cognitive work, not merely cognitive tasks found in science and R\&D. This class of definitions is also used by the International AI Safety Report.\footnote{Bengio (2025).}

  \item \textbf{Milestones focusing on AI's ability to match or exceed humans at all work, including physical work.} Physical capability often serves as a critical AGI threshold because it represents AI's ability to operate in the full human environment. DeepMind's AGI framework currently excludes physical tasks, whilst OpenAI's definition---requiring superiority at `most economically valuable work'---implicitly includes them.\footnote{Google DeepMind (2024) and OpenAI (2025a), respectively.} Popular prediction markets similarly include physical milestones, such as robots assembling scale models.\footnote{For example, Metaculus (2025) asks: `When will the first general AI system be devised, tested, and publicly announced?'.} This matters because much of global value creation depends on embodied work, and even knowledge-based professions such as medicine require physical skills. Progress here would mark AI's expansion from the digital realm into the full spectrum of human activity. This class of definitions places the most stringent demands on the term AGI.
\end{enumerate}

\begin{figure}[htbp]
  \centering
  \includegraphics[width=0.85\textwidth]{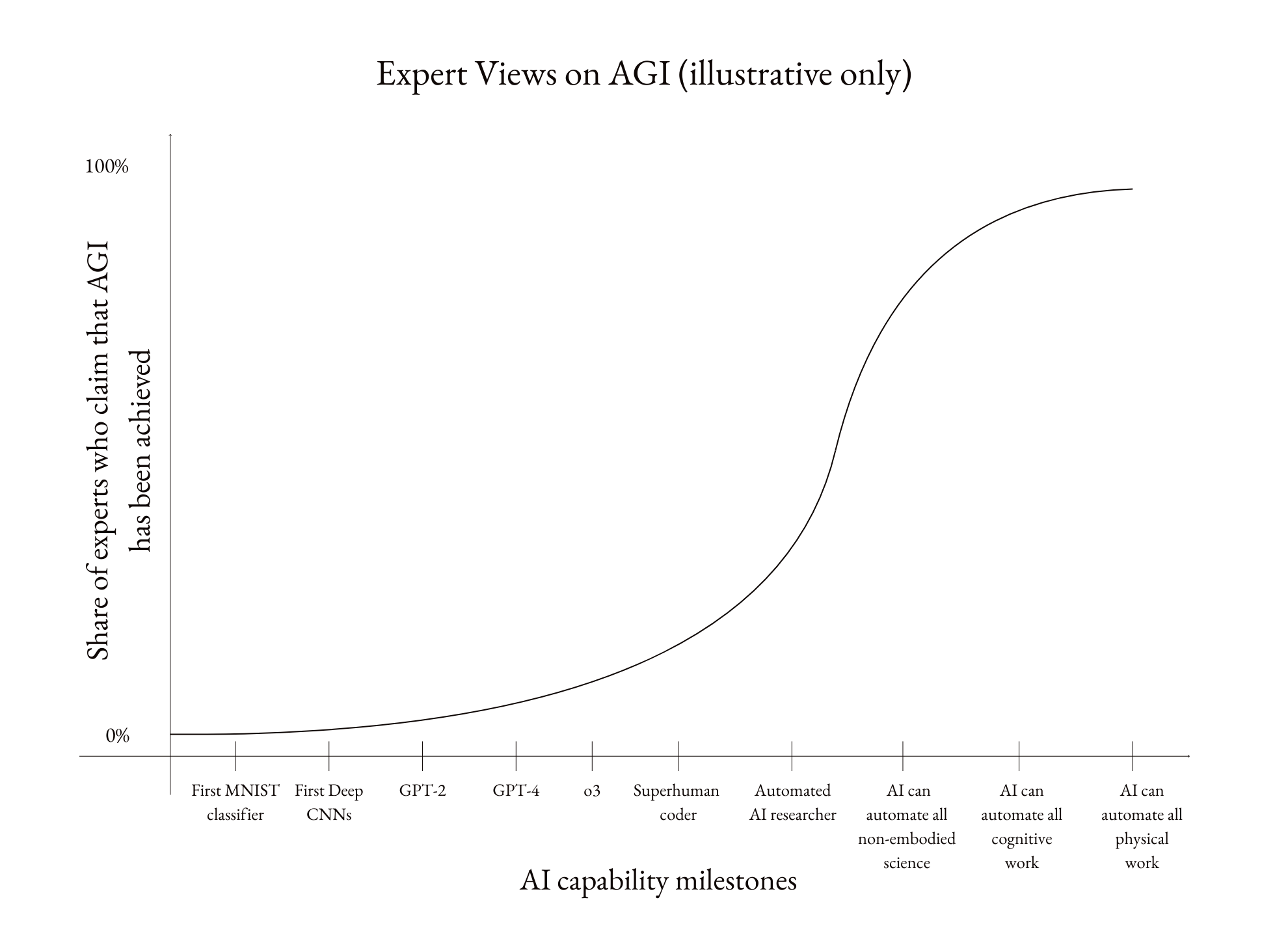}
  \caption{Illustrative development of expert views on whether AGI has arrived. Source: Centre for Future Generations.}
  \label{fig:agi-milestones}
\end{figure}

The categorisation above shows the wide range of possible thresholds for AGI. These four categories of milestones can also be viewed as representing a plausible progression of AI capabilities over the coming years. If progress continues fastest in areas with clear success criteria and feedback loops where data can be easily obtained or generated, we might first witness automation of AI R\&D, followed by broader automation of science and R\&D. Automating most cognitive work may require additional time, until finally AI systems might automate all types of work, including physical tasks through robotics. Figure~\ref{fig:agi-milestones} provides a schematic representation of this possible sequence and illustrates how we can interpret AGI as a spectrum with different milestones. It also demonstrates how experts might continue disagreeing about whether the industry has achieved AGI until AI systems reach broad, superhuman capabilities.

\section{How would AGI affect geopolitics?}
\label{sec:geopolitics}

In this chapter, we investigate the claim that the emergence of AI systems that equal or surpass human performance could affect the very foundations of national competitiveness and lead to a shift in the global balance of power. While the precise characteristics, speed and magnitude of a such a shift remain contested, the literature on the geopolitics of AGI highlights two main pathways by which it could unfold: (1) unprecedented economic growth and technological leverage from large-scale labour automation and accelerated scientific discovery; and (2) shifts in military capabilities that generate a significant, potentially decisive, advantage on the battlefield.

In the economic domain, the broad deployment of AGI could boost economic growth by drastically increasing the amount and productivity of labour in the economy. In a non-AGI-based economy, economic growth is limited by the available amount of human labour.\footnote{Epoch AI (2025a).} If autonomous machines can completely substitute for humans, in that they are able to perform all work that humans can do at a lower cost, then this constraint would be lifted. To give a stylised example: firms built on fleets of digital AI agents operating 24/7 at superhuman speed with minimal human oversight could scale output far beyond human-only organisations.\footnote{Amodei (2024).} Such a shift would plausibly support much faster growth than possible in economies relying solely on human labour.

As a result, countries with rapid or even explosive growth from AGI might be able to outgrow geopolitical rivals without access to AGI. Faster wealth accumulation and compounding returns from automated science and R\&D could, over time, concentrate economic resources and thus geopolitical weight in AGI-powered states. Moreover, possession of AGI amounts to technological leverage: AGI producers could shape other countries' growth by controlling who receives access to their systems and on what terms. In a world of AGI-makers and AGI-takers, where many economies rely on foreign systems for their economic growth, such dependencies could lead to lasting asymmetries in bargaining power and autonomy. It should be noted that these effects may be limited by deliberate policies to slow down automation to prevent AI-driven job loss or tax erosion.\footnote{Korinek (2024).}

AGI-driven growth scenarios, although based on assumptions that are increasingly widely shared among economists and policymakers, come with a significant degree of uncertainty.\footnote{Potlogea \& Ho (2025).} In general, the option space of the potential economic impacts of AGI is large,\footnote{Trammell \& Korinek (2023).} and even opinions about the near-term economic effects of AI on the economy vary widely.\footnote{Korinek \& Suh (2024).} MIT economist Daron Acemoglu predicts excess GDP from AI of just 1.1 to 1.6 per cent over the next 10 years,\footnote{Acemoglu (2024).} while others predict growth rates well in excess of 10 per cent a year over the same time frame.\footnote{Examples include Aschenbrenner (2024a), Kokotajlo et al.\ (2025), Epoch AI (2025a) and MacAskill \& Moorhouse (2025).} A recurring disagreement within the economic literature is whether AI should be understood as a relatively `normal' technology, whose impact is constrained by organisational and institutional frictions, or as a genuinely transformative one capable of driving much faster and more disruptive growth.\footnote{Narayanan \& Kapoor (2025).}

Economic benefits from AGI could also translate into military advantage.\footnote{Black et al.\ (2024).} Recent research suggests that advanced AI could change warfare by improving tasks such as information analysis, knowledge generation, management of complicated actions and systems, and decision making.\footnote{Burdette et al.\ (2025).} It is argued that these improvements could make it easier to deploy massed autonomous systems that would give quantity an edge over quality,\footnote{This is already the case today for some existing uncrewed and crewed platforms, but AGI might reinforce the tendency.} which would enable sophisticated deception campaigns through deployment of large numbers of decoys (a `fog-of-war machine'). Militaries that do not adapt to these changes could be at a serious disadvantage.

Some even argue that the strategic advantages conferred by AGI-based weapons would lead to a significant or even decisive military advantage.\footnote{Mitre \& Predd (2025).} One main argument in favour of this hypothesis is that automated R\&D would lead to a rapid acceleration of military innovation leading to novel `wonderweapons', without much time for enemies to adapt. States with AGI-equipped militaries would then be able to decisively, and perhaps permanently, overpower non-AGI-based militaries. Some authors draw an analogy to the First Gulf War, in which technologically advanced coalition forces completely overwhelmed the Iraqi army---at the time the 4th-largest in the world---within a short period of time and with minimal losses.\footnote{Aschenbrenner (2024b); Zilgalvis (2025).}\textsuperscript{,}\footnote{Critics of this analogy argue that the Gulf War outcome did not solely come about through American technological dominance---it also reflected Iraq's poor training, command structure, morale and international isolation.}

Beyond superior weapon systems, advanced AI could raise overall military and defence industrial productivity. AI-enabled planning, logistics, training and production management could allow states to field more and better equipped forces from the same resource base and to adapt faster in conflict. In the nuclear domain, integrating advanced AI into intelligence, surveillance, reconnaissance, decision support and elements of nuclear command, control and communications could improve situational awareness and the survivability of forces, but might also introduce new vulnerabilities and compress decision times.\footnote{Black et al.\ (2024).} Although most of this analysis concerns advanced AI in general, the effects would likely be amplified by AGI-level systems. AGI development would affect the offence-defence balance, crisis stability and deterrence dynamics, yet current analyses differ on whether the net effect would be stabilising or destabilising.\footnote{Black et al.\ (2024).}

These assessments come with significant uncertainty. In general, most technological innovations do not lead to immediate, discontinuous shifts in military advantage. Some `wonderweapons' arguably have existed; examples might include the tank, stealth technology, the nuclear triad, the Enigma code-breaking machine (`The Bombe'), and early reconnaissance satellites. However, these are historically rare.\footnote{Blanchette et al.\ (unpublished manuscript).} Overall the available evidence suggests that neither transformative economic growth induced by AGI, nor the idea that AGI-based warfare will disrupt the military balance of power, should be accepted as an inevitability.

\subsection{AGI-induced power shifts could lead to global instability}

The literature suggests that AGI-induced shifts in the global balance of power could be accompanied by significant instability. States and sub-state actors with access to AGI capabilities, and the necessary resources to deploy those capabilities at scale, might wield significant influence while non-AGI states might find themselves in a position of dependence and diminished agency.\footnote{Mitre \& Predd (2025).} For some of the latter states, this prospect would be seen as an unacceptable existential threat: there would be no credible guarantee that their core national interests, or their political and physical integrity, would be respected.\footnote{Mueller (2025).} As a result, the path to an AGI-driven world might lead to conflict, as states may choose to engage in preventive action to deny geopolitical rivals a decisive AGI lead, or to protect their own AGI lead from runner-ups.

One form of preventive action is to try to outcompete rivals in developing AGI at all costs, or at least to compete strongly enough to secure a significant share of its future benefits. Pursuing this path could increase AI safety and security risks, since actors may prioritise speed over caution, relax safeguards, compress testing and evaluation, and cut corners.\footnote{Folinsbee (2025).} Another form of preventive action would be to hamper or even destroy or substantially weaken an adversary's AGI project, for example through export controls, or cyberattacks on datacentres. In the most extreme form, if states are convinced that conflict over AGI is inevitable, and that the likelihood of winning such a conflict will only decline in the future, they might consider waging a preventive war, although this has been a historically rare course of action.\footnote{Burdette \& Demelash (2025b).}

The extent to which fears of AGI-driven power shifts generate instability will depend on several factors: the actual and perceived military and economic advantages conferred by AGI; the degree of visibility into one another's capabilities; the design of effective international agreements and verification mechanisms; leaders' assessments of other states' intentions; and the perceived and actual effectiveness of preventive measures. It should be noted that adversarial preventive action is not the only plausible path here. States might not act on supposed strategic imperatives due to risk aversion or other psychological or political factors.\footnote{Mueller (2025).} Moreover, instead of competing or waging war, they might choose the path of diplomacy to seek a cooperative solution.\footnote{Burdette \& Demelash (2025b).}

Some argue that instability around AGI could naturally produce a deterrence regime similar to mutually assured destruction (MAD), in which states threaten to sabotage one another's projects if they suspect efforts to weaponise AGI. In practice this could involve destroying or disabling specialised compute clusters, or corrupting or deleting model weights and training data. Hendrycks, Schmidt \& Wang (2025) describe a version of this scenario as `Mutually Assured AI Malfunction' (MAIM) and suggest it could constitute a stable equilibrium worth maintaining. Critics argue that, while it is worthwhile to think about strategic stability equilibria in AGI worlds, effectively `maiming' a rival's project would be exceedingly difficult, and that key differences with nuclear MAD---including that such institutional theory building could take years or decades and might require near-catastrophes to convince parties to accept mutual vulnerability---may make MAIM more likely to accelerate escalation than to dampen it.\footnote{Rehman, Mueller \& Mazarr (2025).}\textsuperscript{,}\footnote{An important consideration is that AI is a software-centric, dual-use capability, while nuclear weapons are not (even if some nuclear technologies are dual-use).}

Again, none of these dynamics are guaranteed, and the extent to which AGI increases geopolitical instability remains very uncertain. That said, especially given recent geopolitical hardening and increasing competition between the US and China, it would be unwise for Europe to prematurely discard these possibilities.

\subsection{AGI could shift power to individuals or AI systems themselves}

A significant strand of the literature emphasises that shifts of power from AGI may also occur beyond the nation state level. In the absence of effective control mechanisms, AGI capabilities could diffuse to malicious non-state actors (e.g.\ corporate elites, criminal groups or terrorists) and empower them to do significant harm. Firstly, AGI capabilities could lead to novel weapons of mass destruction and widen the circle of actors with access to such weapons. Secondly, humanity could lose control to AGI systems that are unaligned with human values. And thirdly, AGI could lead to a concentration of power in the hands of individuals or small groups of people.

The first scenario becomes apparent when considering how advancements in dual-use AI capabilities have already resulted in new forms of misuse. Text and image generation engines have been repurposed to manipulate public opinion or spread misinformation. Coding capabilities enhance the productivity of software developers, but also enable malicious actors to exploit vulnerabilities in critical infrastructure.\footnote{Bengio et al.\ (2025).} In the future, advanced AI assistants that help manage everyday tasks might also be weaponised for psychological manipulation, subtly influencing individuals' emotions and behaviours---for example to commit fraud at a vast scale or to subvert democratic processes.\footnote{Gabriel et al.\ (2024).}

A dual-use capability of particular concern is the AI-assisted planning and execution of offensive operations, especially in the cyber and CBRN domains.\footnote{Bengio et al.\ (2025).} An example is the emergence of increasingly powerful biological design tools. Current AI systems, such as those developed by Google DeepMind, have already achieved superhuman-level capabilities in modelling protein folding, earning a Nobel Prize and revolutionising pharmaceutical development.\footnote{Desai et al.\ (2024).} However, experts are worried that related tools could be misused to create bioweapons of unprecedented potency, and that they might lower the barrier for novices to design such weapons.\footnote{As highlighted in the International AI Safety Report; Bengio et al.\ (2025).} With AGI, biological design tools could reach even higher levels of sophistication and access, potentially amplifying their potential for widespread harm.

Existing control mechanisms may not be sufficient to keep dual-use AI out of the wrong hands. For example, some security measures designed to limit misuse in open-weight models can be bypassed through targeted fine-tuning.\footnote{Adamson \& Allen (2025); Qi et al.\ (2023).} Closed-weight models offer more protection against this, but cybersecurity measures to prevent weight theft could remain inadequate to ensure non-proliferation.\footnote{Nevo et al.\ (2024).} Without addressing these underlying issues of AI security, there remains a tangible risk that terrorist groups and other malicious actors could become capable of designing, manufacturing and deploying weapons of mass destruction.

The second scenario is rooted in the fact that aligning AI systems to human preferences remains an unsolved scientific problem. This includes both the challenge to operationalise what humans actually want as well as the training of models that reliably pursue such operationalised goals across novel contexts.\footnote{Jank\r{u} et al.\ (2024).} Once considered a merely theoretical concern, empirical evidence for misalignment is accumulating as the science of model evaluations advances. For example, recent research by Anthropic has demonstrated that agentic models sometimes resort to blackmailing and leaking of sensitive information when they perceive their core goals to be under threat.\footnote{Anthropic (2025a).} There is also increasingly salient evidence of AI systems pursuing goals covertly: recent work by Apollo Research demonstrates that AI systems are able to take sophisticated measures to prevent their own disempowerment, for example by creating hidden backups of themselves.\footnote{Apollo Research (2025).}

While misaligned AI systems can already cause limited harm today, problems could multiply with powerful, agentic AGI systems that autonomously pursue long-term goals and command significant resources.\footnote{Taylor (2025).} Some argue that an accidentally unaligned system may optimise for objectives that are actively harmful to human interests and could seize control over humanity (`active loss of control').\footnote{As highlighted in the International AI Safety Report; Bengio et al.\ (2025).} A more subtle, yet related possibility is humans becoming increasingly dependent on AI systems, up to the point at which they lose ability for oversight, leading to gradual disempowerment (`passive loss of control').\footnote{Kulveit et al.\ (2025); also as highlighted in Bengio et al.\ (2025).}

The third scenario---that of concentration of power---comes in at least two variations. In one variant, concentration of power emerges as a byproduct of AGI-driven automation.\footnote{Korinek \& Suh (2024).} As increasingly capable systems substitute for human labour across most tasks, wages may collapse while returns to capital surge, shifting income from workers to the owners of compute, models and data. Wide joblessness and societal disruption follow, as human labour retains little economic bargaining power, and as the world moves towards a durable concentration of both wealth and political influence in the hands of those few who control the AGI capital stock.\footnote{Korinek (2024).} Over time, such an imbalance could hollow out the material foundations of political participation, making it increasingly difficult for democratic institutions to meaningfully constrain those who control AGI.\footnote{Bullock, Hammond \& Krier (2025).}

Another variant of power concentration is that of a group of people deliberately taking power. This idea of an AGI-enabled coup has become the focus of serious analysis only recently and is currently based on mostly theoretical arguments.\footnote{Davidson, Finnveden \& Hadshar (2025).} The idea is that AGI systems could make it much easier for small groups of people to seize power, as they would avoid some of the failure modes of conventional human-powered coup attempts. Unlike humans, AGI systems could be singularly and unwaveringly loyal to institutional leaders, and this loyalty could be made secret and hard to detect. If control over such loyal and powerful AGI `sleeper agents' is concentrated in the hands of a few actors, such as AI company CEOs or small groups of senior government officials, this might enable a swift overpowering of conventional institutional checks and balances, paving the way to autocracy.

The potential for AGI-enabled power concentration shows that aligning AGI to human interests is necessary, but not sufficient to avoid a bad outcome. Once AGI is aligned to human interests, those who control it need to be aligned with the wider interests of society and constrained by democratic checks and balances. These checks and balances likely need to be much more robust than policies addressing present-day wealth inequality and conventional mechanisms protecting institutions from human-enabled power grabs.

Safeguards against all three risk scenarios described in this section are under urgent development, but it is not guaranteed that they will be sufficient.\footnote{Bengio et al.\ (2025).} Model evaluations are making rapid strides, but the science of evaluations is still in relative infancy.\footnote{Apollo Research (2024).} Approaches to AI safety such as neural network interpretability, scalable oversight and AI control offer promising avenues for progress in terms of understanding and containing AI systems, but none is currently advanced enough to offer sufficient assurance for more advanced models.\footnote{Ngo, Chan \& Mindermann (2023).} Technical and legal measures against AGI-enabled coups remain theoretical,\footnote{Davidson, Finnveden \& Hadshar (2025).} and economic policy responses to AGI-based wealth concentration are only in their infancy.\footnote{Korinek (2024).} A European approach to AGI preparedness should reflect that these crucial gaps in technical AI safety and governance remain unresolved, and may not be resolved in the short-to-medium term without a major coordinated effort.

\subsection{Expected power shifts are driving competing superpowers to treat AI and AGI as a strategic priority}

Although extreme, AGI-induced power shifts have not yet materialised, recent political developments suggest that governments increasingly view advanced AI as a key determinant for economic and military competitiveness. Across administrations, the US government has viewed leadership in AI as a strategic imperative. Under the Biden administration, the US National Security Memorandum on AI stated that `ceding the United States' technological edge would not only greatly harm American national security, but it would also undermine United States foreign policy objectives and erode safety, human rights, and democratic norms worldwide.'\footnote{White House (2024).} The Trump administration voices similar priorities, albeit with more martial rhetoric: its recent US AI Action Plan opens with the statement that `the US is in a race to achieve global dominance in artificial intelligence.'\footnote{White House (2025).}

In accordance with this rhetoric, the US is taking ambitious steps to entrench its current leadership in AI capabilities, and to prevent diffusion of the most advanced technologies to geopolitical rivals. Ben Buchanan, former AI advisor to President Biden, has stated publicly that the US government is taking the possibility of AGI seriously and is taking active steps to prepare.\footnote{Klein (2025).} While the Trump administration has reversed some of Biden's AI policies, it continues to prioritise strengthening domestic AI capabilities, by taking steps to encourage the construction of datacentres, semiconductor manufacturing facilities and energy infrastructure.\footnote{White House (2025).} To prevent foreign diffusion, the Trump administration has so far continued semiconductor export controls in order to deny China access to lithography tools and the most advanced AI chips, although export of a scaled-down Nvidia Blackwell chip named the B30A and removal of some export control provisions appear to be on the table.\footnote{Adamson et al.\ (2025).}

Parallel but distinct developments can be observed in China, the second AI superpower. The Chinese government has stated that it wants to become the global leader in artificial intelligence by 2030, aiming to turn AI into a \$100bn industry and to create more than \$1 trillion of AI-generated value in other industries.\footnote{Chan et al.\ (2025).} In addition to existing investments by the private sector, the government is using a variety of tools to boost its domestic AI ecosystem: supporting the development of domestic AI chips as a response to export controls, supporting AI research at state laboratories and building large datacentres to rival US computing capacity.\footnote{Olcott et al.\ (2025); Chan et al.\ (2025).} A protocol from a Politburo discussion chaired by Xi Jinping in April 2023 also revealed that the Chinese leadership is considering AGI at the highest political level.\footnote{Tao (2025).}

This apparent equivalence in strategic awareness and policy ambition has led some observers to the diagnosis that the US and China are embroiled in a race to AGI.\footnote{Vlassov (2025).} This is true in the sense that both American and Chinese AI companies are stating AGI as a goal and that the US government currently frames itself as being in an AI race.\footnote{Milmo (2024); Tao (2025).} However, it should also be noted that state rhetoric in China is less focused on winning a race and more focused on economic benefits from AI.\footnote{Chan et al.\ (2025).} Critics have thus argued that the notion of an A(G)I race should be closely examined so as not to turn into a self-fulfilling prophecy.\footnote{\'{O} h\'{E}igeartaigh (2025).} Although the increasing salience of the potential stakes of AGI will increase incentives to race, whether to race or not remains a policy choice and not a law of nature.

Outside the two superpowers, a growing group of non-EU AI middle powers is treating AI with comparable strategic seriousness and searching for room to manoeuvre. The UK wants to position itself as an `AI maker, not an AI taker', has forged a \pounds{}31bn investment partnership with US hyperscalers,\footnote{UK Government (2025b).} AI companies and chip manufacturers,\footnote{Kirkwood (2025).} and continues to provide generous financial support to its AI Security Institute.\footnote{UK Government (2025a).} The United Arab Emirates (UAE) is viewing compute as an opportunity to diversify away from oil: its state-backed AI investment vehicle MGX is one of the main funders of the US Stargate project.\footnote{Abbott (2025).} Moreover, several UAE leaders, including UAE president Sheikh Tahnoun bin Zayed Al Nahyan, are reported to be `true believers' in AGI.\footnote{Allen et al.\ (2025).} Singapore, meanwhile, is emerging as a leader in AI safety diplomacy: its Singapore Consensus on Global AI Safety Research Priorities brought together a broad coalition of AI researchers and articulated a shared global roadmap for AI safety research.\footnote{Singapore Conference on AI (2025).}

In conclusion, the literature suggests that concerns around geopolitical shifts of power and instability are no longer theoretical. Instead, the contours of a new geopolitical order are emerging: superpowers are accelerating to AGI, middle powers are scrambling for agency, and safety and security risks have moved beyond the political agenda. Europe needs to orient itself to this new reality and act strategically if it wants to defend its prosperity and values. The next chapter evaluates whether Europe's current level of preparation is adequate to rise to the challenge of AGI.

\section{Is Europe sufficiently prepared for a fast transition to AGI?}
\label{sec:europe-prepared}

Assessing whether Europe is prepared to navigate a potential transition to AGI requires a structured examination of the changes and requirements that AGI's emergence would impose on governments, economies and societies. Drawing on literature exploring strategic foresight,\footnote{An important example includes Kent's book on `Strategic Intelligence for American World Policy' (1949), which argued that policy makers operating without accurate assessment of adversary capabilities, technological trajectories and emerging threats will systematically make poor decisions regardless of resources or intentions. Work by Buchanan (2020) makes similar arguments for AI, where capabilities evolve rapidly, implications spread across multiple domains, and information asymmetries are common.} societal transitions\footnote{For example, technological transition studies such as Geels (2002) and Genus \& Coles (2008).} and geopolitical competition,\footnote{For example, scholars have identified several dimensions of competitive positioning that determine geopolitical outcomes: control over critical inputs, as in Farrell \& Newman's (2019) research on `weaponized interdependence' or ability to set standards and rules as argued by Bradford (2019). For AI competition, these often translate into questions about compute infrastructure, talent, data, capital and the ability to influence global governance.} we evaluate AGI preparedness through three dimensions: strategic awareness in government, competitive positioning and leverage, and the robustness of current policy strategies. These feed into a government's ability to: (1) perceive and interpret frontier AI developments accurately and rapidly; (2) command resources and leverage points relevant to AI development and governance; and (3) formulate and execute coherent plans on AI that align resources with objectives under conditions of uncertainty.

\textbf{Dimension 1: Strategic awareness} assesses whether European institutions possess the organisational capacity to accurately perceive, interpret and communicate information about frontier AI developments and their implications. Operationally, this includes:
\begin{itemize}
  \item \emph{Tracking trends in frontier AI capabilities:} Monitoring frontier capabilities and their probable trajectories; assessing the potential and limitations of emerging AI paradigms, such as, most recently, AI agents, reasoning models and decentralised training.
  \item \emph{Studying the societal impacts of capabilities:} Understanding how large-scale deployment of AI systems could affect society; for example, evaluating agentic AI's effects on the labour market and persuasive AI's effects on democratic processes and interpersonal relationships.
  \item \emph{Understanding the national security implications of capabilities:} Assessing potential large-scale threats from AI capabilities; for example, the ability of models to conduct large-scale cyberattacks, to create bioweapons, and the risk of loss of human control.
  \item \emph{Collecting strategic and competitive intelligence:} Assessing a country's standing relative to geopolitical competitors and evaluating others' technological capability and access to assets in the AI value chain.
\end{itemize}

\textbf{Dimension 2: Competitive positioning and leverage} assesses whether Europe has assets in the AI value chain that allow it influence AI development and ensure reliable access to frontier capabilities; we examine both Europe's competitive position across key input factors and access to `chokepoints' where European actors hold disproportionate influence.\footnote{This approach is inspired by the theory of `weaponised interdependence' as described by Farrell \& Newman (2019).} For competitive positioning, we assess:
\begin{itemize}
  \item \emph{Model capabilities:} Where do current European frontier AI models rank relative to US and Chinese competitors, and is the gap projected to widen or narrow?
  \item \emph{Compute infrastructure:} What share of global AI computing capacity does Europe control, and is it sufficient for competitive frontier development?
  \item \emph{Complementary inputs to development:} Does Europe possess adequate quality and volume of data, capital, energy and talent---each necessary for successful AI development?
  \item \emph{Diffusion capacity:} Can Europe translate AI capabilities into adoption that leads to meaningful productivity gains, whether in government, the military, the private sector and wider society?
\end{itemize}

For leverage, we examine:
\begin{itemize}
  \item \emph{Supply chain positioning:} Does Europe control critical nodes in the AI supply chain (e.g.\ semiconductors, rare materials, manufacturing equipment) that could provide bargaining power in AI competition?
  \item \emph{Market power:} Can Europe leverage its Single Market credibly to extract governance commitments and capability guarantees from foreign AI providers?
\end{itemize}

Crucially, we assess whether these strategic assets could be effectively deployed under conditions of geopolitical stress and when they would be subject to specific countermeasures.

\textbf{Dimension 3: Robustness of current policy strategies} assesses whether Europe has formulated ambitious, sufficiently integrated and well-resourced plans that set out realistic objectives. We analyse:
\begin{itemize}
  \item \emph{Strategic clarity:} Has Europe articulated explicit goals for AGI development, access, and governance, with clear prioritisation between these?
  \item \emph{Use of resources:} Are committed resources commensurate with stated ambitions and the scale of the challenge?
  \item \emph{Integration mechanisms:} Do coordination mechanisms exist to align efforts across fragmented institutional landscapes (EU vs Member States, multiple DGs, civil vs defence domains)?
  \item \emph{Adaptation capacity:} Can strategies be rapidly modified if AI trajectories diverge from planning assumptions, or are they locked into long decision cycles?
\end{itemize}

We assess this dimension by analysing major EU AI initiatives (AI Continent Action Plan, Apply AI Strategy, Frontier AI Initiative, InvestAI, etc.) against the criteria listed above. We do not claim to be exhaustive and acknowledge that the enforcement and implementation of these strategies carries perhaps equal weight as their mere articulation.

The three dimensions correspond to distinct failure modes that could undermine European AGI preparedness. Without adequate strategic awareness, Europe might fail to anticipate when critical capability thresholds are reached, misallocate resources (e.g.\ in compute infrastructure), or be left unready for systemic risks such as mass unemployment and nation state-led cyber attacks. In addition, without competitive positioning and sufficient leverage in the global value chain, Europe might formulate strategies based on accurate awareness of technological trends but lack the capacity to execute them, thereby becoming a `rule-taker' whose preferences are sidelined in global AI governance. Finally, without a coherent strategy Europe's efforts might fragment, resulting in an inability to make difficult but necessary trade-offs. By focusing on these three core dimensions, this framework allows for a structured diagnosis of potential weaknesses and blind spots in Europe's strategic stance.

\subsection{Dimension 1: Strategic awareness}

\subsubsection{Europe has limited strategic awareness of frontier AI progress}

Here we examine whether European governments maintain a comprehensive, real-time understanding of the landscape and trajectory of frontier AI capabilities, and their potential societal and geopolitical implications. For example, AGI-focused labour market policies need to be informed by a detailed understanding of which sectors AGI automation is likely to occur in first, military planners need to take into account how AI is likely to shift the offence-defence balance and the drivers of strategic advantage, and geopolitical strategy requires an understanding of which countries and actors have access to which AI capabilities and assets along the supply chain. As established above, strategic awareness constitutes a necessary precondition for effective action---without accurate assessment of technological trajectories and their implications, even well-resourced actors may systematically make poor decisions. Across all domains, decision makers need to understand at which speed and in which directions the `jagged frontier' of capabilities is likely to evolve. Moreover, pan-European and national authorities need to act with a clearly defined division of labour to avoid oversight or duplication.

In Europe, a particularly advanced example of institutional capacity is the UK AI Security Institute (UK AISI). Founded in 2023, it describes its mission as `equip[ing] governments with a scientific understanding of the risks posed by advanced AI.'\footnote{As shown on UK AISI's website; UK AISI (2025).} As an ambitious `start-up in government' with wide-ranging exemptions from regular civil service hiring and salary structures, and endowed with \pounds{}100m\footnote{In the recent 2025 spending review, the UK government (2025b) pledged an additional \pounds{}240m (\texteuro{}273m) to UK AISI.} (${\sim}$\texteuro{}113m) of initial funding, it managed to assemble a team of world-leading AI researchers within the first three months of its inception.\footnote{UK Government (2023).} A recent addition to UK AISI is the `strategic awareness team', which is directly tasked with `conduct[ing] deep dives on critical uncertainties regarding AI trajectories and impacts, track[ing] key indicators and warning signals, and produc[ing] briefs for key government decision makers.'\footnote{An open role to join UK AISI's strategic awareness team as a Research Scientist highlights this; see UK AISI (2024).}

UK AISI is a part of the wider Department for Science, Innovation and Technology (DSIT), where it is complemented by a number of additional units of expertise, for example relating to frontier AI regulation, international AI policy and AI innovation policy.\footnote{Internal RAND analysis.} Notably, DSIT has a Sovereign AI Unit, endowed with up to \pounds{}500m (${\sim}$\texteuro{}570m), and tasked with fostering UK sovereignty in a world with advanced AI.\footnote{UK Government (2025b).} There are also dedicated AI units in the Foreign, Commonwealth \& Development Office, the UK Ministry of Defence and the Cabinet Office. Although official numbers are not available, informed estimates suggest that across both UK AISI and DSIT at least 300 FTEs (and likely more) are working on various aspects of strategic awareness and AI policy.\footnote{UK government source.} Arguably, the strategic awareness generated by these institutions has enabled the UK to lead in a number of AI policy domains; the country hosted the first AI summit in Bletchley, has wide-ranging testing access to leading US AI models, and has negotiated a UK--US tech agreement involving an investment of \pounds{}31bn (\texteuro{}35bn) in UK AI infrastructure.\footnote{UK Government (2025c).}

State capacity for strategic awareness in continental Europe has increased, but remains less advanced. The leading centre for expertise on AI in the EU is the EU AI Office (EU AIO). While the EU AIO has attracted an impressive cohort of prominent AI researchers and scientists, is endowed with a globally unique legal mandate to gather information on and regulate frontier AI companies, and has the power to convene EU Member States on topics of advanced AI, it remains limited in its mandate and capabilities. Its starting budget---at \texteuro{}46m---is less than half that of UK AISI, despite its much greater jurisdiction, and its more powerful enforcement mandate. Its team is not large enough to act as a `strategic awareness function' for the EU institutions, much less for the individual Member State governments, and recruiting more people is challenging as its hiring scales and payment process are less flexible than those of UK AISI, hampering its ability to attract technical talent fast enough.\footnote{Schaefer (2025).} Finally, unlike UK AISI and the US Center for AI Standards and Innovation (US CAISI), the EU AIO does not have a mandate to work on matters of defence and national security, as these remain Member State responsibilities.\footnote{The EU AI Office is structured into 6 units and 2 advisors, which reflects its mandate: the `Excellence in AI and Robotics' unit, the `Regulation and Compliance' unit, the `AI Safety' unit, the `AI Innovation and Policy Coordination' unit, the `AI for Societal Good' unit, the `AI in Health and Life Science' unit, the Lead Scientific Advisor and the Advisor for International Affairs. EU AI Office (2025).}

Key EU Member States are also making progress on strategic awareness capacity, but institutional set-ups are not as ambitious as they could be given the gravity of a potential AGI challenge. France is the only Member State with an AI Safety Institute: instead of creating a new legal structure for an AISI, France chose to create a consortium of existing institutes for evaluation and security, which were collectively branded the `National Institute for the Evaluation and Security of Artificial Intelligence' (INESIA). INESIA is supervised jointly by the National Security Secretariat and the Directorate General of Enterprises at the Ministry of Economics and Finance. Unlike the EU AI Office, it has an explicit mandate to work on national security topics, but critics argue that INESIA does not have the same budget stability and `start-up in government' flexibility as UK AISI, and that its decentralised institutional structure may lead to siloing and inter-institutional rivalries.\footnote{Barbier, Martinet \& Segerie (2025).}

Germany, the largest country in the EU, does not have an AISI or an equivalent strategic awareness function in government. There are, however, centres of AI expertise across government, such as in AI-focused units in the Federal Chancellery, the Ministry for Digital and State Modernisation, the Ministry of Economic Affairs, the Ministry for Research, Technology and Space, the Ministry of Defence, the Foreign Office, and agencies such as the German Federal Office for Information Security. Unlike France, Germany does not participate in important information-sharing fora such as the International Network of AI Safety Institutes, and unlike in the UK there has been no concerted effort in Germany to recruit top-tier technical AI talent into government to strengthen strategic awareness. Consequently, many of the most qualified German AI scientists who could become part of a strategic awareness function continue to work in Silicon Valley or in foreign AISIs.

Overall, while continental Europe has made strides in strategic awareness capacity on AI, most prominently in the form of the EU AI Office and the French INESIA, its institutions do not appear well-resourced and focused enough to be on par with ambitious initiatives such as the combined UK AISI/DSIT cluster of expertise. As a consequence, European policy makers risk being less well informed about key developments in AI, including the potential implications of AGI. This could result in worse policy design as well as decisions that are lacking the urgency and ambition necessary to compete with rival powers and to effectively mitigate downside risks. Fortunately, addressing this shortcoming may be more straightforward compared to other policy challenges associated with AI. Expanding state capacity for strategic awareness is less a matter of missing talent or expertise than of political will and prioritisation to embed existing expertise directly in government.

\subsection{Dimension 2: Competitive positioning and leverage}

Here we evaluate Europe's competitive positioning and leverage across key inputs for AI development and governance. We divide this assessment into two components: firstly, Europe's competitive position across core input factors (models, compute, data, capital, energy, talent and diffusion capacity), and secondly Europe's leverage points in the AI supply chain and global market that could amplify its global influence beyond raw `inputs'.

We conclude that the EU lags significantly behind the US and China across most indicators of AI development. This includes frontier AI capabilities, where European models trail US and Chinese leaders by 6--12 months\footnote{Artificial Analysis (2025b).}; computational infrastructure, with Europe controlling only 5 per cent of global AI compute capacity\footnote{As presented in Pilz et al.\ (2025) data in `Trends in AI Supercomputers'.}; data access, where the shift towards synthetic data disadvantages compute-constrained Europe, while stricter privacy regulations constrain data reuse\footnote{Lakshmanan (2024).}; capital availability, with EU start-ups attracting just 6 per cent of global AI venture funding\footnote{Martens (2024).}; energy costs and availability, with Europe facing structurally higher electricity prices and permitting bottlenecks\footnote{Heussaff (2024); European Commission (2025c); SolarPower Europe (2024).}; and talent retention, with Europe training AI researchers but losing many of the best to better-funded American companies.\footnote{Pal (2024).} Europe also lags in AI diffusion capacity---the ability to translate frontier capabilities into productivity gains across the economy.

\subsubsection{Models: European AI companies lag behind the frontier by 6--12 months}

\begin{figure}[htbp]
  \centering
  \includegraphics[width=0.85\textwidth]{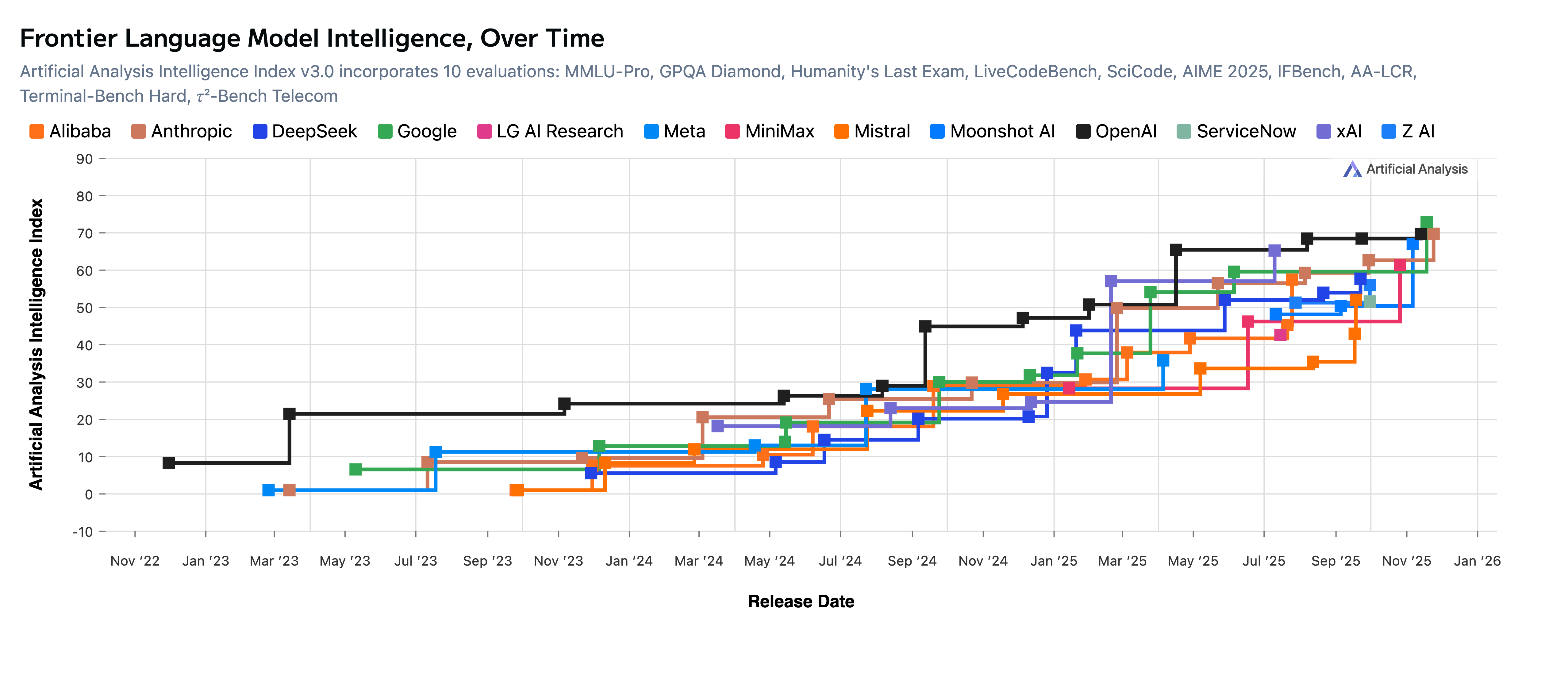}
  \caption{Benchmark performance of European frontier models in international comparison. Source: Artificial Analysis (2025a).}
  \label{fig:eu-models}
\end{figure}

European AI companies struggle to match the performance of leading models. European AI, led by Mistral's innovations in multilingual reasoning and open-source development, remains behind the US and Chinese models in terms of both performance and cost efficiency. Mistral's most advanced model, Magistral Medium, scores 75.3 per cent on MMLU-Pro---a widely used benchmark that tests broad knowledge and reasoning across 57 subjects ranging from math and physics to law and history---compared to DeepSeek V3.1's 85.1 per cent and Claude 4.1 Opus' 88 per cent.\footnote{Artificial Analysis (2025a).} At the same time, Mistral's models are more costly, at \$2--5 per million tokens versus DeepSeek's \$0.50--\$2 and gpt-oss-120B (high)'s \$0.15--\$0.60 pricing. Latest estimates signal that Mistral's Magistral Medium lags behind the best US and Chinese models by 6--12 months (see Figure~\ref{fig:eu-models}).\footnote{Artificial Analysis (2025b).}

Europe is simultaneously outmatched in terms of overall numbers of frontier models. Excluding Google DeepMind (which is based in the UK and part of a US company), Europe produced only three notable models in 2024 compared to 40 from the US and 15 from China, highlighting a significant gap in model production volume, which serves as an indicator of the depth of the AI ecosystem.\footnote{Stanford University (2025). Epoch AI (2025h) defines `notable model' as `meeting any of the following criteria: (i) state-of-the-art improvement on a recognized benchmark; (ii) highly cited (over 1000 citations); (iii) historical relevance; (iv) significant use'.} With very few European developers at frontier scale, a lock-in to a non-European AI tech stack looks increasingly plausible. The cloud computing market also demonstrates this dynamic, where early, aggressive scaling by US hyperscalers such as AWS and Microsoft Azure created powerful economic moats and high switching costs for European customers.\footnote{Srivathsan, Sorel \& Sachdeva (2024).}

The gap with China is also widening, with Chinese models getting closer to the capabilities of US offerings. DeepSeek's R1 achieved near-parity with leading US models when it came out in May 2025, lagging behind then-leading Claude Sonnet 4 by only 2 percentage points on GPQA Diamond, which tracks performance on expert-level questions in science, mathematics and technical domains.\footnote{Epoch AI (2024).} The broader trend echoes similar patterns: performance gaps on major benchmarks such as MMLU narrowed from 17.5 percentage points in 2023 to just 0.3 points by early 2025.\footnote{Stanford University (2025), part `3. The gap between Chinese and US models closes.'} Industry analysts now estimate Chinese models lag behind US frontiers by only 3--6 months,\footnote{Insikt Group (2025).} an improvement from the 1+ year gap that was estimated in 2022.\footnote{Artificial Analysis (2025b).} Figure~\ref{fig:eu-models} compares the benchmarks used to assess these models.\footnote{Incorporates 8 evaluations: MMLU-Pro, GPQA Diamond, Humanity's Last Exam, LiveCodeBench, SciCode, AIME, IFBench and AA-LCR.}

While lags of only a few months behind the frontier might not seem especially pressing, they already translate into massive commercial disparities: OpenAI generated \$10bn in annualised recurring revenue by June 2025,\footnote{Capoot \& Rooney (2025).} while Mistral had revenue of \$30m in 2024, with projections of \$60m for 2025\footnote{D'Souza (2025).}---more than a hundredfold difference. This suggests a winner-takes-most dynamic in AI markets, with superior capabilities driving exponentially higher customer adoption, pricing power and market share. With such market dynamics, it is an open question whether European champions will be able to continue financing their AI development. Moreover, the gap becomes even more consequential as leading models begin automating large parts of the AI development process itself, triggering feedback loops that could compound advantage for the leader. The usage of AI to automate AI development makes the capabilities gap more likely to widen rather than close in the future, even when rivals are currently only months behind.\footnote{Jank\r{u} et al.\ (2025).}

\subsubsection{Data: European AI companies are constrained by compute and data regulation}

Data constitutes a foundational input for AI development, serving as the raw material from which models learn patterns, develop capabilities and adapt to specific use cases. The quality, quantity and diversity of training data determine a model's performance, generalisation ability and real-world applicability. However, the data requirements for frontier AI are undergoing a change that affects competitive dynamics between regions.

Until recently, frontier models relied primarily on scraping vast quantities of internet text, images and code for `pre-training'---the stage where an AI system learns general patterns before being taught specific tasks.\footnote{The legality of training AI on copyrighted material without permission remains contested, with ongoing litigation in multiple jurisdictions. Even where courts have found such use legally permissible under fair use doctrine, the ethics of training on creators' work without consent or compensation remains disputed.} After pre-training, low-skilled human labellers were needed to ensure AI systems' outputs were useful, ethical and in line with the developer's guidelines. As detailed in Chapter~\ref{sec:how-far-away}'s discussion of how reinforcement learning might circumvent the `data wall', AI developers have since developed computationally intensive post-training and runtime techniques that rely heavily on synthetic data---data produced and filtered by AI systems themselves.\footnote{Epoch AI (2025f, 2025g).} Looking ahead, the most valuable human contributions to AI training may evolve from large-scale data production and low-skilled labelling to creating highly specialised datasets---particularly in domains requiring tacit knowledge. Similarly, data collection may shift from mostly passive gathering to the construction of interactive reinforcement learning environments where AI agents can explore and learn autonomously. These changes mean that in addition to the volume of human-generated content, expert data and computing power have become a core determinant in the quest for training data.\footnote{Wexler (2025).}

Europe is at a disadvantage when it comes to synthetic data generation. Limited access to computing capacity compared to the US and China may constrain the ability of European researchers to generate and refine the massive quantities of synthetic data now driving AI progress. Europe also faces disadvantages in the realm of non-synthetic training data. US firms control large quantities of user-generated content, interaction logs and cloud data that can be repurposed for AI development under existing terms of service.\footnote{OECD (2025).} The US also lacks comprehensive federal data protection law, and broad interpretations of `fair use' in copyright have allowed AI companies to train on vast quantities of scraped content with limited legal friction---though this permissiveness is now being tested in litigation such as \emph{The New York Times v.\ OpenAI}.\footnote{Roth (2023).} China combines large domestic user bases with strong state coordination and mandated data-sharing among domestic companies, enabling both scale and centralisation. China's 2021 data protection laws nominally impose restrictions comparable to GDPR,\footnote{Zhu (2022).} but enforcement is asymmetric: domestic AI champions benefit from state-mandated data-sharing arrangements and lighter scrutiny, while foreign competitors face stricter compliance burdens.

Europe's General Data Protection Regulation (GDPR) poses additional constraints. Purpose limitation requirements mean data collected for one service cannot be freely repurposed for AI training without establishing a new legal basis.\footnote{GDPRhub (2025), Article 5(1)(b)).} Reliance on `legitimate interest' for AI training thus faces increasing regulatory scepticism. This was illustrated by Meta's June 2024 decision to pause EU AI training following complaints by the Austrian privacy group NOYB and intervention by the Irish Data Protection Commission. Data minimisation principles complicate foundation model development, which benefits from maximal data ingestion. The right to erasure also creates a hard technical challenge for models that have already learned from subsequently deleted data.\footnote{Moreno (2025).} These structural factors constrain the manoeuvring space of European actors in data-intensive AI development.

The above factors help to explain why promises of European data advantages have not (yet) materialised. Earlier EU strategies, including the 2020 EU Data Strategy\footnote{European Commission (2020).} and 2021 Fostering a European Approach to AI,\footnote{European Commission (2021).} anticipated that Europe's industry-specific knowledge and high-quality domain data would provide competitive advantage. However, this data remains fragmented across sectors and largely underutilised.\footnote{Batur \& Peeters (2021).} The Commission's recent AI Continent Action Plan\footnote{European Commission (2025a).} represents a more promising approach, aiming to expand data availability for EU developers by streamlining regulatory access. By establishing Data Labs and Common European Data Spaces, it aims to create curated, high-quality and interoperable datasets in strategic areas such as health, manufacturing and energy. Yet even with improved data access, Europe will require substantial AI expertise to transform passive datasets into the interactive reinforcement learning environments that frontier AI development increasingly demands.\footnote{Patel \& Kourabi (2025).}

\subsubsection{Compute: Europe hosts only 5 per cent of the world's AI computing infrastructure}

The development and deployment of advanced AI systems depends on access to massive computational infrastructure. Training frontier models currently requires large clusters of specialised chips running in centralised facilities costing billions of euros and often consuming as much power as small cities. Meanwhile inference---the process of deploying trained models to generate output---requires less raw power per datacentre site but represents an ongoing operational cost that scales with usage. Companies and governments that control AI infrastructure not only derive economic benefits from AI development and adoption, but also have more control over who can build and use AI systems and related services.

\begin{figure}[htbp]
  \centering
  \includegraphics[width=0.85\textwidth]{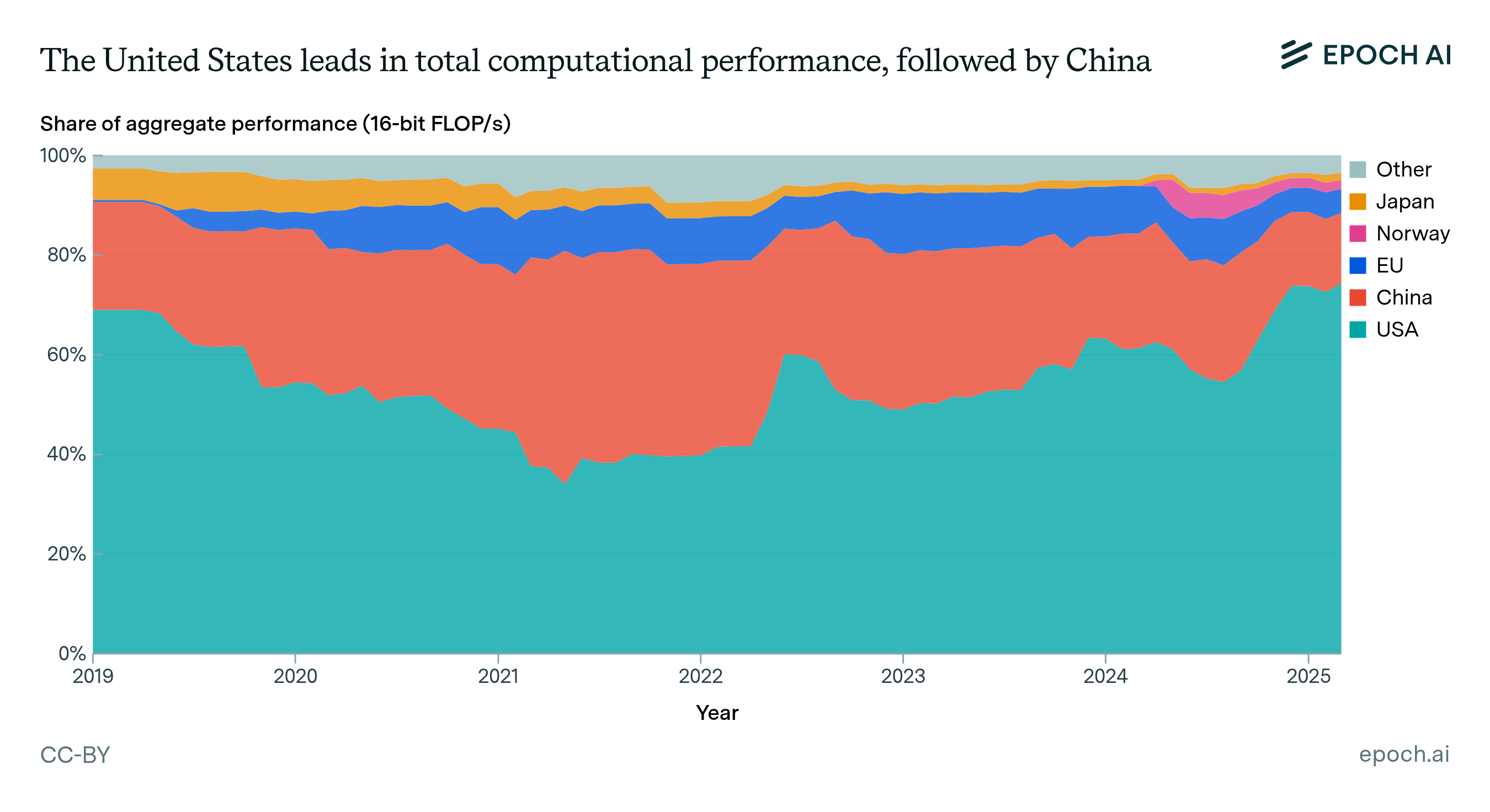}
  \caption{Europe's share of global aggregate computational performance (in 16-bit FLOP/s). Source: Pilz, Rahman et al.\ (2025).}
  \label{fig:compute}
\end{figure}

Europe is far behind the US and China in terms of housing onshore AI infrastructure (see Figure~\ref{fig:compute}). The US has the world's largest share---about 75 per cent---of global aggregate computational performance for AI, followed by China with 14 per cent and the EU at nearly 5 per cent.\footnote{Pilz et al.\ (2025) have a dataset on `Trends in AI supercomputers' that gathers this data. These figures require cautious interpretation, as the study captures only 10--20 per cent of AI supercomputers globally, with uneven coverage between private and public infrastructure, and focuses exclusively on supercomputers rather than overall compute performance. This means that government supercomputers are over-captured in the data and the overall AI compute landscape may appear different in practice.} This disparity may be partially explained by Europe's higher electricity prices,\footnote{Patel (2024a).} long permitting timelines and lack of excess grid capacity.\footnote{Kothuri et al.\ (2025).} Some improvements are on the horizon. Planned AI Gigafactories will bring investment of \texteuro{}20bn, while the Cloud and AI Development Act aims to triple the EU's datacentre stock in the next 5--7 years.\footnote{European Commission (2025a).} Meanwhile, France plans investments in AI infrastructure of \texteuro{}109bn, framing it as a response to the US Stargate project, and has announced the establishment of Europe's largest AI campus in Paris.\footnote{Piquard (2025).} Even so, this still might not lead to a relative advantage, since the US may also triple its datacentre capacity between 2023 and 2027.\footnote{Patel (2024a).} As a result, the gap between Europe and its primary competitors will likely stay constant or widen further.

Leading US hyperscalers and AI companies are demonstrating continued commitment to build the next generation of datacentres underpinning America's global AI leadership. In 2025 alone, US hyperscalers announced plans for investments in AI datacentres worth almost \texteuro{}300bn.\footnote{Subin (2025).} In addition, the Stargate project schedules investments of \$500bn (\texteuro{}455bn) over the next 5 years to construct the supercomputer facilities powering OpenAI's future frontier models,\footnote{Murgia, Hammond \& Kinder (2025).} Oracle has signed a \$300bn contract with OpenAI to provide them with computing resources over 5 years starting in 2027,\footnote{Plumb (2025).} and NVIDIA has invested another \$100bn (\texteuro{}87bn) in OpenAI to enable them to buy more NVIDIA chips\footnote{OpenAI (2025c).} supported by the public funding of \$138bn (\texteuro{}119bn).\footnote{Parasnis (2025).}

Global competition for datacentre construction has intensified recently, with the Middle East becoming a key destination for vast AI compute hubs.\footnote{Guardian Staff and Agencies (2025).} For example, the UAE is planning a huge `AI campus' in Abu Dhabi---touted as the largest outside the US---and could be permitted to import up to 500,000 cutting-edge AI chips each year, starting in 2025. The agreement also commits the UAE to investing in US data centres of comparable scale and to aligning its security standards with those of the US.

The widening AI infrastructure gap may soon become a liability for Europe's economic competitiveness and capacity for innovation.\footnote{Shamir \& Negele (2025).} Insufficient access to onshore European compute directly limits the ability of European firms to develop, test and deploy advanced AI systems, from experimental research to large-scale frontier models. While accessing cloud services from foreign hyperscalers offers an immediate and effective solution, this approach has some limitations. Firstly, it reinforces the network effects cementing US-based providers' dominance, stifling European alternatives. Secondly, geographically distant foreign datacentres would fail to deliver the ultra-low latency required by some critical industrial applications---advanced manufacturing processes, autonomous vehicle systems and real-time robotics---which depend on inference speeds where even millisecond delays can compromise safety and efficiency. Thirdly, it exposes European industry to the increasingly constrained supply of cloud services.\footnote{Srivathsan, Sorel \& Sachdeva (2024).} As global demand for AI training and inference surges, reliable access becomes a competitive differentiator, and over-reliance on external providers leaves European innovation vulnerable to prioritisation decisions made elsewhere.

Beyond the economic implications, a dependency on foreign AI infrastructure threatens Europe's security and strategic autonomy.\footnote{Hawkins, Lehdonvirta, \& Wu (2025).} A dearth of reliable, onshore compute capacity jeopardises the resilience of Europe's most critical infrastructure and industries, potentially leaving essential services vulnerable to supply chain disruptions or geopolitical coercion. Moreover, reliance on foreign datacentres also cedes control over vital security standards, from hardware-enabled safeguards to physical security, undermining Europe's ability to protect sensitive government and commercial data as well as frontier model weights according to its own requirements.\footnote{Nevo et al.\ (2024).} Without sufficient onshore AI infrastructure, Europe's ability to act independently in a geopolitical crisis---or alternatively, its ability to shape global AI governance in line with its values---may be constrained.

\subsubsection{Capital: only a small share of global AI funding is going to EU-based start-ups}

The computational power required to train today's frontier models translates into a massive demand for capital. Acquiring and networking hundreds of thousands of GPUs and running them for months at a time to complete a single pre-training run requires a large upfront expense. Unlike inference, which distributes costs over time, training is a capital-intensive, high-risk R\&D expenditure that must be incurred before any revenue is generated. In this environment, access to patient and risk-tolerant capital becomes a prerequisite to being and staying at the frontier. This tight link between extreme compute costs and large-scale finance makes Europe's fragmented financial system a structural weakness in frontier AI competition.

In the first half of 2024 more than \$35bn was invested into AI start-ups worldwide, yet only around 6 per cent of that funding went to EU-based start-ups.\footnote{Martens (2024).} At the same time, Europe struggles to mobilise patient institutional capital that can wait years for returns. As the Draghi Report highlights, EU pension assets in 2022 amounted to only about 32 per cent of GDP, compared to 142 per cent in the US and 100 per cent in the UK, while private pension savings are heavily concentrated in a few Member States such as the Netherlands, Denmark and Sweden.\footnote{Draghi (2024).} In the meantime, training costs for frontier AI models are surging, growing at a rate of 2.4 times per year since 2016.\footnote{Cottier (2024).} Assuming that rate stays constant, the largest training runs, without factoring in much larger fixed datacentre expenditures, will cost more than a billion dollars by 2027---a scale of investment that privileges capital-rich, risk-taking incumbents.\footnote{Cottier (2024).} The Draghi Report already warns that 30 per cent of European unicorns relocated abroad between 2008 and 2021, primarily to the US, to access deeper funding pools.\footnote{Draghi (2024).} Without stronger cross-border capital integration, Europe risks pushing aspiring frontier AI champions overseas.

\subsubsection{Energy: electricity prices are significantly higher than in the US}

AI infrastructure demands vast electrical power. As AI capabilities and adoption grow quickly, energy use is becoming a development bottleneck as important as access to advanced chips or capital. Recent research estimates that global AI datacentres could require an additional 10 gigawatts of power capacity in 2025, almost twice the entire power capacity of Ireland.\footnote{Estimates included in research by Pilz, Mahmood \& Heim (2025). Ireland's system demand---the electricity production needed to meet national consumption---is at 5.23 gigawatts in 2025, as the electricity Transmission System Operator (TSO) for Ireland shows; EirGrid (2025).} Assuming that chip supply can keep up with demand, AI datacentres could need around 68 gigawatts in total by 2027, slightly more than the total power capacity of Poland today.\footnote{Estimates included in research by Pilz, Mahmood \& Heim (2025). Poland's grand system capacity is at 64.3 gigawatts in 2025, as per the EU electricity transparency hub; Entsoe (2025).} This rapid increase in demand, driven by both model training and inference, is turning power availability into a core dimension of global AI competition.

Europe's energy system has features that are particularly poorly matched to this. The continent faces structurally higher energy prices: in 2023, industrial electricity prices in the EU were 158 per cent higher than in the US, and industrial gas prices were 345 per cent higher.\footnote{Heussaff (2024).} This largely reflects Europe's reliance on imported fossil fuels, a dependency exacerbated by Russia's invasion of Ukraine. Certain countries, such as Germany, have also been sceptical of nuclear power as a solution to long-term energy needs. By contrast, the US is a net energy exporter. Europe's energy security also rests on decarbonisation-focused policy frameworks, such as the European Green Deal, which aims for long-term autonomy through renewables but may impose higher short-term costs than continued fossil fuel use.

\begin{figure}[htbp]
  \centering
  \includegraphics[width=0.85\textwidth]{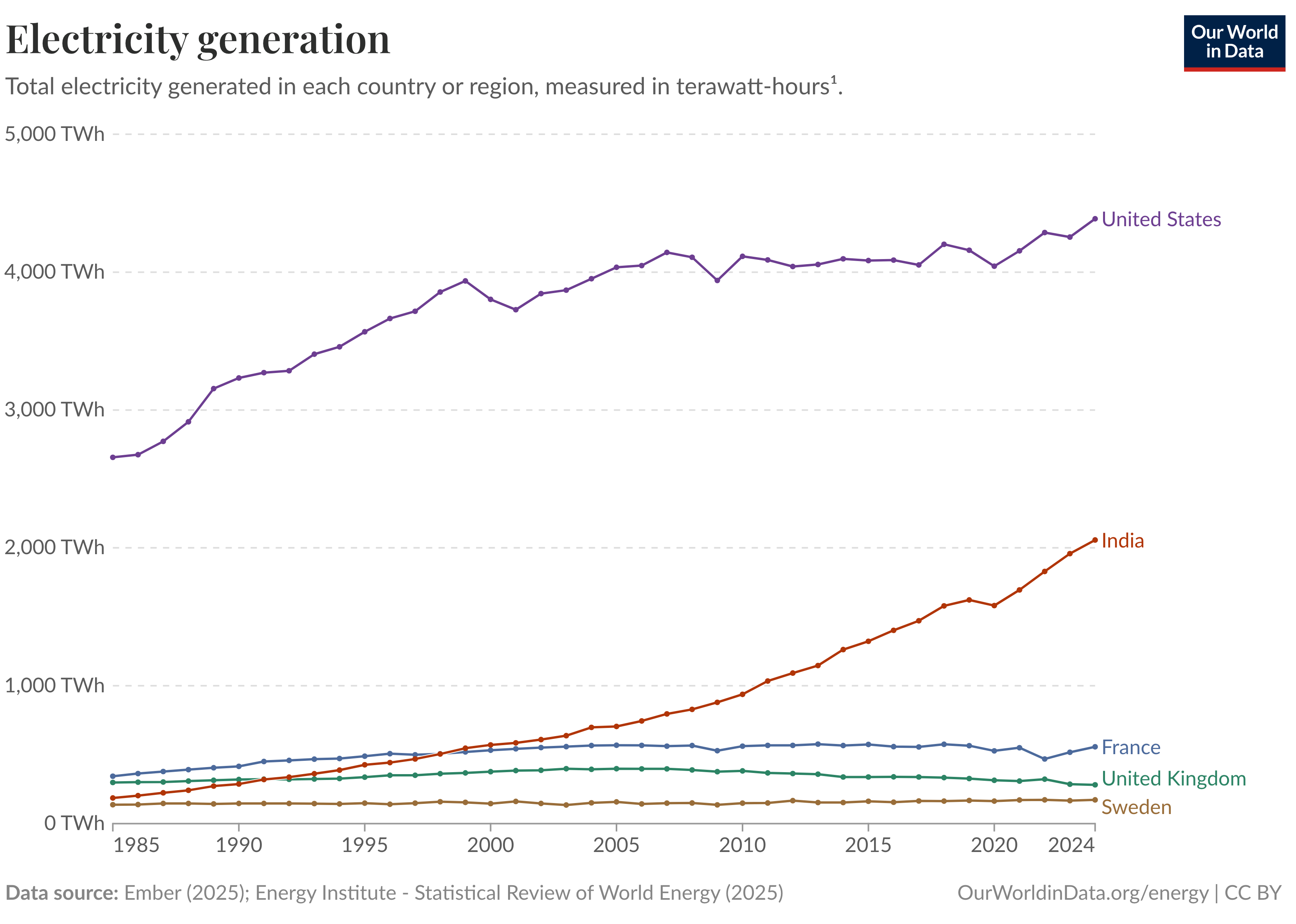}
  \caption{Total electricity generated in each country or region (in terawatt-hours). Source: Our World in Data (2025).}
  \label{fig:energy}
\end{figure}

In addition, unlike the mostly integrated continental grids of the US and China, Europe's energy system remains a collection of separate national grids.\footnote{Entsoe (2025a).} While interconnected, they face cross-border bottlenecks and divergent national regulations, which makes it harder to provide the large-scale, predictable power capacity that massive datacentre clusters require.\footnote{Draghi (2024).} Permitting is a further constraint: average approval times for new datacentres stand at about 48 months, and EU-level planning reforms have often been slowed by the late transposition of permitting rules in several Member States.\footnote{European Commission (2025c); SolarPower Europe (2024).}

In contrast, other AI powers enjoy stronger energy security. The US benefits from large domestic energy resources, especially natural gas, which keeps power relatively cheap and reliable. This allows US hyperscalers to plan and build large onshore datacentres with confidence about long-term energy availability and cost.\footnote{Rueda et al.\ (2025).} To manage grid congestion and intermittency, US firms also invest abroad; for example, Microsoft and OpenAI's Stargate project taps Middle Eastern markets with abundant, low cost energy to complement domestic supply.\footnote{Cornish \& Acton (2025).} China, meanwhile, uses its state-driven model to implement a centrally planned energy strategy, directing massive investments into generation capacity from renewables to nuclear at a speed and scale unmatched in Europe and the US (see Figure~\ref{fig:energy}).\footnote{Roytburg (2025).} Its `Eastern Data, Western Computing' strategy, for example, strategically couples newly constructed datacentres with nearby large-scale power supply.\footnote{Mu Miao (2024).}

Energy capacity will remain a critical input for AI development and adoption, and this matters well beyond model training. Modern AI systems consume power every time they process tokens for text, images or reasoning tasks.\footnote{Salvator (2025).} As capabilities advance and inference becomes more compute intensive and more important for productivity, the energy cost per unit of useful output will increasingly matter. This is especially true if reasoning-heavy models and autonomous AI agents see broad adoption in critical sectors. In such a world, regions with cheaper and more abundant electricity can run more AI for the same cost, giving them an edge in developing and deploying transformative systems.\footnote{Heim \& Lin (2025).}

\subsubsection{Elite talent concentration: little of the top AI talent works in Europe}

Talent is a critical input for Europe's ability to build frontier AI models, given the importance of algorithmic efficiency and the complexity of very large training runs. China hosts the deepest pool of AI talent, with around 47 per cent of the world's top tier AI researchers.\footnote{Defined by MacroPolo (2024) Global AI Talent Tracker as researchers who publish in highly selective venues such as NeurIPS, especially those whose work is chosen for oral presentation; used as a proxy for the world's most elite AI researchers.} The US is the main talent magnet, with about 60 per cent of top AI institutions and a majority of global top AI talent (much of it from Europe) working in US organisations in 2022. In comparison, about 12 per cent of top AI talent worked in Chinese institutions, 8 per cent in continental Europe and 8 per cent in the UK.\footnote{MacroPolo (2024).}

In terms of quantity, the EU is good at training talent but poor at retaining it. Many leading researchers move to the US, while India has become a key source of AI professionals for Europe. For example, 28 per cent of Ireland's, 15 per cent of Sweden's, and 13 per cent of Germany's AI workforce completed their undergraduate degrees in India.\footnote{Pal (2024).} Recent trends show some EU countries gaining AI workers overall\footnote{Stanford University (2025).} and recent political developments in the US may be reversing traditional talent flows.\footnote{Reif (2025).} As many as 75 per cent of American scientists who responded to a 2025 Nature poll on the new US administration's policies indicated that they are contemplating relocating abroad.\footnote{Witze (2025).}

On talent quality, the EU faces an acute shortage of advanced AI engineers, with fewer than 0.63 qualified candidates per vacancy for these positions.\footnote{Pal, Schneider \& Nurski (2025).} Most of the EU's AI talent consists of `technical professionals in software and data roles using basic AI methods'---for those, the supply exceeds demand. However, the shortfall in `advanced AI engineers and researchers developing cutting-edge models or other advanced AI techniques' limits the EU's ability to reach the technological frontier.

Europe's AI workforce also suffers from fragmentation. Several EU countries have more AI experts per capita than the US or China, but this talent is spread across many small hubs.\footnote{Pal, Marino Lazzaroni \& Mendoza (2024).} In the US, AI professionals cluster in a few large centres such as San Jose, San Francisco and Seattle. In the EU, only Amsterdam, Berlin and Paris rank among the top 25 global AI hubs, whilst London ranks 8th globally.\footnote{Pal, Marino Lazzaroni \& Mendoza (2024).} This limits the agglomeration effects and innovation spillovers that drive breakthrough AI development in more concentrated ecosystems like Silicon Valley.

Taken together, Europe trains a large AI workforce but loses many of its strongest researchers to the US, and much of the remaining talent is spread across many small hubs rather than concentrated in major tech centres.\footnote{Pal, Schneider \& Nurski (2025).} As a result, Europe currently lacks the dense pool of advanced expertise needed to compete in building frontier AI models.

\subsubsection{AI diffusion: European adoption of AI is a mixed story, but likely somewhat behind the US}

Europe's ability to capture economic value from frontier AI depends on how well it can diffuse the technology across its economy. Ding's work\footnote{Ding (2024).} on general purpose technologies shows that diffusion capacity often determines who captures gains and can matter more than invention itself. Policy priorities increasingly reflect this focus on diffusion: both the US AI Action Plan\footnote{White House (2025).} and the EU's Apply AI Strategy\footnote{European Commission (2025b).} emphasise adoption, with the latter mobilizing \texteuro{}1bn across 11 strategic sectors.

\begin{figure}[htbp]
  \centering
  \includegraphics[width=0.85\textwidth]{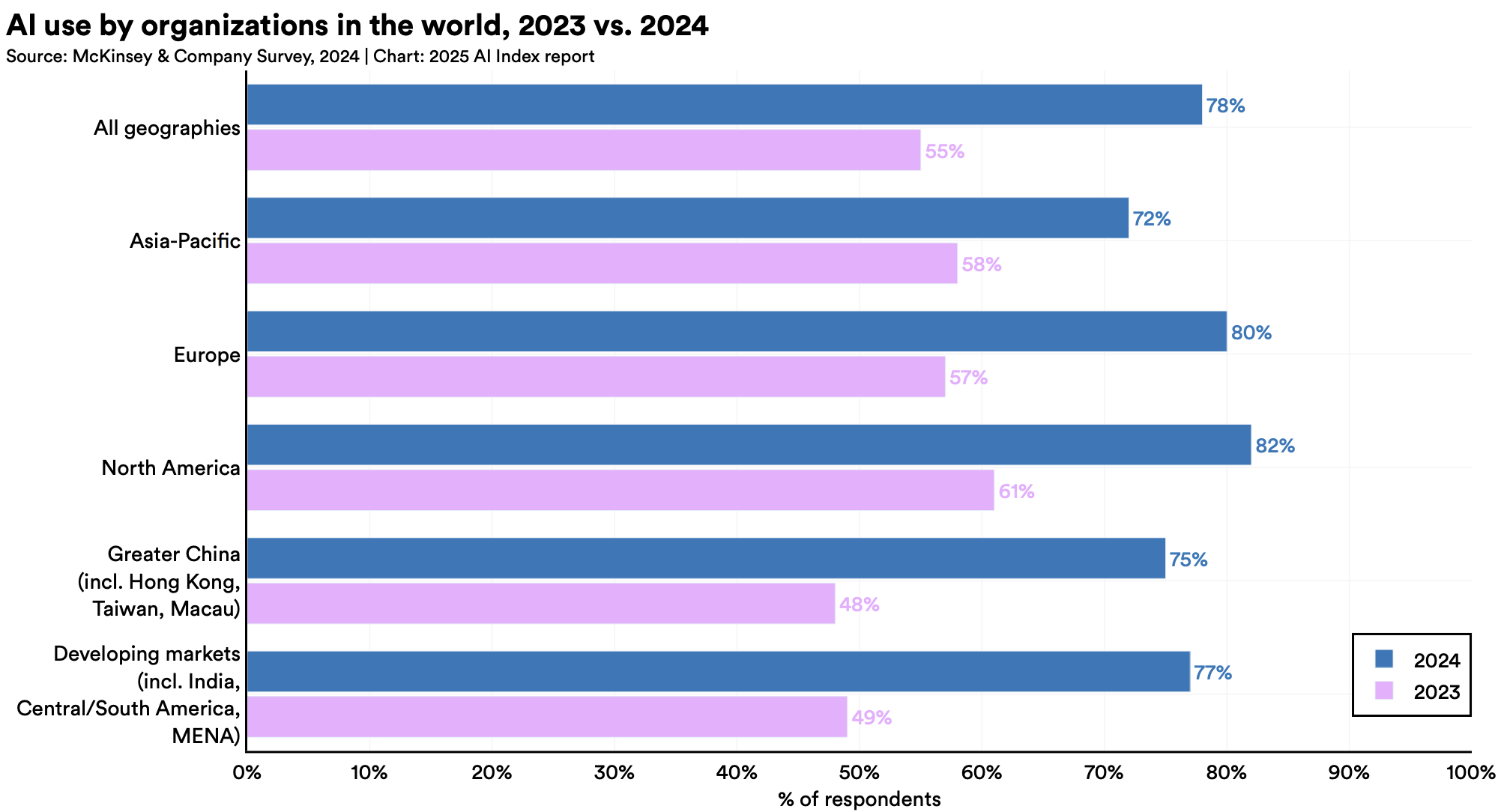}
  \caption{AI use by organisations in the world, 2023 vs.\ 2024. Source: Stanford University (2025).}
  \label{fig:ai-adoption}
\end{figure}

Current metrics of AI adoption paint an ambiguous picture. According to Eurostat,\footnote{Eurostat (2025).} only 13.5 per cent of EU firms reported using AI in 2024, with strong variation by firm size and country: 41 per cent of large firms versus 11 per cent of small firms, and 24 per cent in Denmark versus 4 per cent in Romania. International comparisons are hard because of different survey methods, but these numbers alone do not reveal a large transatlantic gap.\footnote{Kergroach \& H\'{e}ritier (2025).} One US survey recorded about 8 per cent of firms using AI to produce goods or services within a two week window in April 2025.\footnote{Census Bureau (2025).} A McKinsey survey cited by the Stanford AI Index finds only a two percentage point gap in broad AI adoption between North America and Europe, with 82 per cent of respondents in North America and 80 per cent in Europe reporting use of AI in at least one business function (see Figure~\ref{fig:ai-adoption}).\footnote{Stanford University (2025).} Beyond these surveys, data comparing adoption rates between regions is limited. Anthropic's AI Usage Index is one notable exception, suggesting that usage rates in the US are clearly ahead of countries in the EU, even those with high adoption rates such as France and Germany.\footnote{Appel et al.\ (2025).} That said, this metric only covers one provider's users, and others, like OpenAI, have not released regional data.

However, current adoption rates tell only part of the story. The key question is whether Europe's economic and institutional environment supports deep, economy wide integration of AI. On paper, Europe's economy appears to be well placed for this. With services accounting for about 65 per cent of GDP, AI agents and other `digital workers' could automate or augment many cognitive and information processing tasks.\footnote{European Union (2023); Denain, Ho \& Sevilla (2025).} In practice, structural frictions such as weak commercialisation pathways may limit how far AI capabilities translate into productivity growth and even a successful diffusion strategy could remain dependent on foreign inputs for models, compute and energy.

A recent commentary by Ding highlights several levers for building diffusion capacity\footnote{Ding (2025).}: enlarging the pool of AI implementers beyond elite researchers through training and certification programmes, creating diffusion institutions that support small and medium-sized firms, ensuring interoperable standards, and providing reliable access to cloud infrastructure and sectoral data. The EU's Apply AI Strategy\footnote{European Commission (2025b).} addresses part of this agenda. It aims to support small and medium-sized firms through about 250 Digital Innovation Hubs and to launch sectoral flagship projects backed by funding. Its main emphasis, however, is on deployment support, rather than on deeper structural reform.

\subsubsection{Europe's main strategic levers provide limited influence}

Europe holds two major assets: a critical role in the global supply chain for advanced semiconductors, most notably Dutch firm ASML's monopoly on extreme-ultraviolet (EUV) lithography equipment, and the EU's market power as the world's second-largest economic area. In theory, these assets could serve as bargaining chips to secure access to frontier AI capabilities and shape AI governance beyond the AI Act. In practice, however, their value is limited by structural dependencies, institutional fragmentation and the risk of retaliation, which may be decisive in periods of geopolitical turbulence. Even so, they remain important sources of leverage for Europe if used strategically.

\paragraph{Semiconductor supply chain: strong positioning, but conversion to influence is not straightforward.}

\begin{figure}[htbp]
  \centering
  \includegraphics[width=0.85\textwidth]{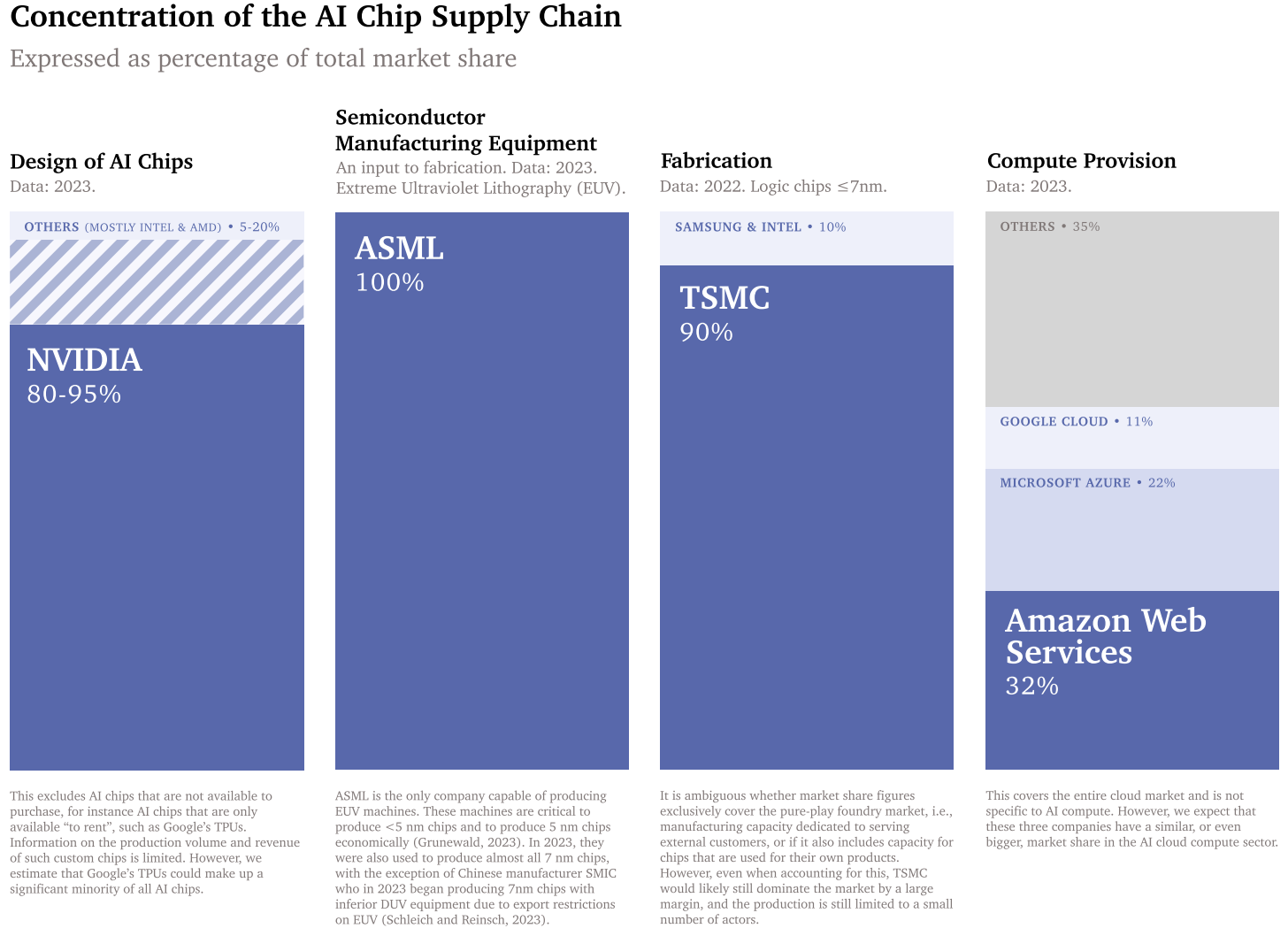}
  \caption{Concentration of the AI chip supply chain, expressed as a percentage of total market share. Source: Heim et al.\ (2024).}
  \label{fig:supply-chain}
\end{figure}

The semiconductor supply chain is the foundation of frontier AI: it is a highly specialised, globally distributed ecosystem in which a few firms control critical chokepoints (see Figure~\ref{fig:supply-chain}). NVIDIA dominates the design of AI chips\footnote{NVIDIA has about 92 per cent market share of the add-in-board GPU market; Farooque (2025).}; Taiwan's TSMC fabricates roughly 90 per cent of the world's most advanced processors\footnote{Sacks \& Segal (2025).}; and South Korea leads in high-bandwidth memory (HBM).\footnote{Yang \& Cherney (2025).} A crucial part of the supply chain is manufacturing equipment, with European firms playing an outsized role. The Dutch firm ASML, supported by specialist suppliers such as the German optics manufacturer Zeiss, has a monopoly on the EUV machines that are essential for printing the smallest, most advanced patterns for cutting-edge chips.\footnote{Clark (2021).} ASML is also the dominant producer of less advanced deep-ultraviolet (DUV) lithography machines,\footnote{Hijink (2024).} but does not hold a complete monopoly since Chinese firms have started to develop their own, albeit less advanced, DUV machines.\footnote{Wentz \& Lin (2025).}

ASML is sometimes regarded as the single most important chokepoint in this chain.\footnote{Lugt (2024).} Precisely quantifying ASML's lead in EUV technology is difficult, but experts regard it as the technology where China is furthest behind and facing the greatest difficulty catching up.\footnote{Grunewald (2023).} Recreating just the laser in an EUV system requires identifying and assembling 457,329 unique parts, which helps explain why these machines are sometimes described as among the most complex objects humans have ever built.\footnote{TRUMPF (2025); Satariano (2025).} Earlier attempts by Nikon and Canon to develop EUV tools failed despite multibillion dollar investments.\footnote{Hijink (2024).} Today, the primary firm reported to be pursuing a comparable EUV capability is China's SiCarrier, operating under the internal code name `Mount Everest'.\footnote{Olcott and Wu (2025).} Nearer-term Chinese efforts to reduce reliance on Western technology focus on DUV tools. Shanghai-based start-up Yuliangsheng has produced a DUV machine now on trial with SMIC, China's leading foundry, and Shanghai Micro Electronics Equipment (SMEE) also produces less advanced DUV systems.\footnote{Shilov (2024); Olcott and Wu (2025).}

Since every advanced AI chip in the world, with the partial exception of Huawei's recent Ascend 910C, is produced using EUV lithography, the ability to control access to EUV machine production and maintenance translates directly into geopolitical power for the Netherlands and the EU. This was illustrated when pressure from the US administration led the Dutch government to block EUV exports to China and later to tighten controls on some DUV tools, moves that significantly slowed China's progress in advanced chipmaking.\footnote{Satariano (2025).}

This position creates strategic leverage for Europe, but using it is not straightforward. The European Union, through the Netherlands and a small group of partners, can influence who has access to the most advanced lithography equipment and thus who can build the compute that powers frontier AI. In theory, the EU could harden controls on China beyond the current EUV ban, for example by restricting DUV exports. In a hypothetical transatlantic dispute, it could even seek to delay equipment or deny on-site maintenance for partners such as TSMC, risking an escalatory violation of the US Foreign Direct Product Rule.\footnote{The FDPR is a US export regulation that extends American jurisdiction to foreign-made items if they are a `direct product' of specified US technology or software or produced by equipment derived from such technology. Because most advanced SME contains US-origin chips or is made with US tools, the FDPR subjects the vast majority of foreign-produced advanced chips and SME to US control, regardless of where they are made. The US has previously leveraged this extraterritorial authority to close loopholes that allow for circumvention via foreign subsidiaries and to compel allies to adopt equivalent controls by offering exemptions. For more details, see Allen \& Goldston (2025).} We stress that these are stylised extreme scenarios rather than policy recommendations, and that each would face major legal, political and practical constraints, as well as serious risks.

Several constraints limit Europe's room for manoeuvre. Firstly, ASML depends heavily on US technology and expertise, notably through its Cymer laser division and substantial US workforce, which exposes it to the US' export control policy.\footnote{Hijink (2024).} Secondly, using ASML as leverage against the US rather than China would collide with Europe's broader dependence on the US for security, cloud services and energy, creating a kind of mutual deterrence. Thirdly, using this leverage more aggressively against China risks strong economic retaliation and could accelerate China's state-funded drive for self-sufficiency.\footnote{Self-sufficiency is already a dominant goal of Chinese semiconductor policy and some may argue that the Chinese ecosystem could not push for much greater acceleration at this time.} Huawei is reportedly to have produced only about 200,000 advanced AI chips in 2025, a fraction of the roughly one million US-produced chips Chinese firms imported in 2024. However, China's leading foundry, SMIC, is already using multi-patterning techniques with existing DUV machines to produce 7nm and potentially 5nm chips,\footnote{Olcott \& Wu (2025).} efforts clearly aimed at closing the technology gap over the long term.\footnote{Olcott \& Wu (2025).} Fourthly, European institutional fragmentation adds another constraint. Despite EU dual use rules and the current EUV export ban to China, export controls remain mainly a national competence with limited EU-level oversight, which creates barriers to rapid and coordinated action. The Pegasus spyware controversy, in which banned tools still received export licences, illustrates how fragmented enforcement can undermine common policy.\footnote{European Parliament (2023).}

How Europe manages its supply chain position is therefore a defining test of its strategic autonomy. The EU, and in particular the Netherlands, is often caught between US pressure for tighter export control alignment and the threat of Chinese economic retaliation.\footnote{Lugt (2024).} For European policy makers, the challenge is to deploy its leverage in a measured, strategic way to deter economic coercion and to shape a global governance regime that protects European interests.

\paragraph{Market power: the Brussels Effect in AI might not last forever.}

Accounting for about 17 per cent of global GDP, the EU's Single Market is one of its main sources of geopolitical leverage in AI.\footnote{IMF (2025).} With roughly 440 million consumers and a GDP of about \texteuro{}18 trillion, the EU is currently the largest trading partner of the US.\footnote{European Commission (2022).} This economic weight provides two related forms of market power: direct leverage to negotiate terms of access to frontier AI capabilities and AI infrastructure, and indirect normative influence through regulatory standard setting, often described as the Brussels Effect.\footnote{Bradford (2019).} These tools have limits and depend on the future strength and attractiveness of the Single Market itself.

\begin{figure}[htbp]
  \centering
  \includegraphics[width=0.6\textwidth]{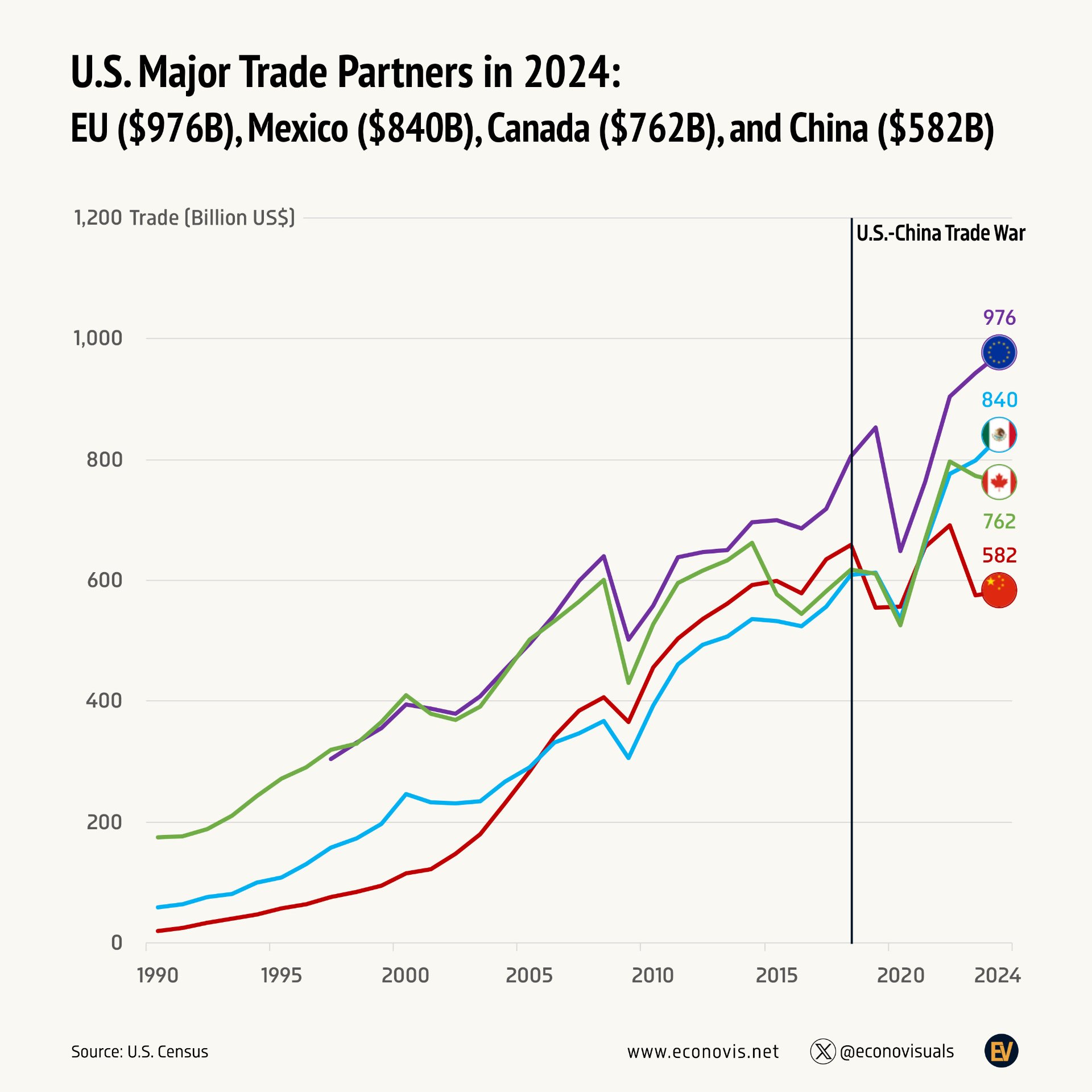}
  \caption{The major trade partners of the US in 2024: the EU (\$976bn total trade), Mexico (\$840bn), Canada (\$762bn) and China (\$582bn). Source: Econovis (2025).}
  \label{fig:trade}
\end{figure}

The EU has long used market access to obtain commitments and shape external behavior.\footnote{Damro (2012).} Switzerland's partial access to the Single Market for goods, air transport and public procurement was granted only on condition that it accept free movement of people.\footnote{Wintour (2016).} In trade policy, the revised EU--Mercosur agreement makes effective implementation of the Paris Agreement an essential element, allowing the EU to suspend trade benefits if climate commitments are seriously violated.\footnote{T\"{a}htinen (2024).} Similar approaches could in principle be applied to AI. The EU could, for example, seek specific conditions on access to frontier models, such as capability assurances or commitments on compute capacity. As training costs rise, with some estimates projecting hundreds of billions of dollars for a single run by 2030 if scaling continues,\footnote{Owen (2025).} US AI firms will need reliable access to foreign markets to recoup their investments. Recent US policy points in the same direction: the US AI Action Plan presents global diffusion of US AI capabilities as a core strategic interest, which suggests that access to the EU Single Market remains commercially and strategically valuable for US firms seeking worldwide reach.\footnote{White House (2025).}

This negotiating position comes with important caveats. Europe's access to leading systems may increasingly depend on how far AI becomes securitised. The Biden administration's now-revoked diffusion framework,\footnote{Bureau of Industry \& Security (2025).} which imposed quantitative semiconductor restrictions on 18 EU Member States and controls on access to closed model weights,\footnote{Heim (2025).} showed that even between NATO allies, access can be constrained when technologies are seen as strategically critical. European users currently enjoy wide access to leading AI models, but this is not guaranteed; as AI systems approach more transformative capabilities, the calculus around diffusion may shift further. Europe is therefore well advised to treat AI access as an issue for active diplomacy and market-based leverage, rather than relying solely on alliance ties.

Beyond deliberate negotiation, the EU exerts influence through the Brussels Effect\footnote{Bradford (2019).}: the tendency of multinational firms to adopt EU rules globally to avoid the cost and complexity of maintaining different standards for different markets. In AI, this effect is particularly relevant for the development of foundation models. The cost of building and maintaining separate, regionally compliant versions encourages firms to align with the strictest standard as their global baseline.\footnote{Siegmann \& Anderljung (2022).} This is already visible in practice. Most leading AI companies, including OpenAI, Google, Microsoft, Amazon, Anthropic, Cohere, IBM and Mistral, have voluntarily signed the EU's GPAI Code of Practice.\footnote{European Commission (2025h).}

Taken together, Europe can convert market access into concrete assurances and can export workable standards when regulatory convergence lowers firms' compliance costs. Maintaining this leverage, however, requires keeping the Single Market attractive and providing predictable compliance pathways. If the regulatory environment is seen as unstable or overly fragmented, firms are more likely to segment features by region, and the EU's market power in AI will be used only selectively rather than as a consistent strategic asset.

\subsection{Dimension 3: Robustness of current policy strategies}

The third dimension of our preparedness framework asks whether Europe has coherent, well-resourced strategies with realistic objectives for its position in AGI geopolitics. The EU has made substantial progress in recent years, and new policy initiatives signal rising ambition. On the governance side, the AI Act creates the world's first comprehensive regulatory framework for AI, and the Code of Practice for general-purpose AI sets a global benchmark for targeted, risk-based compliance.\footnote{European Commission (2025g).} Major AI companies, including OpenAI, Google, Microsoft, Anthropic and xAI, have committed to implement the Code's provisions on transparency and copyright, systemic risk assessment and cybersecurity. This shows that Europe can set proportionate standards that shape global AI development even without hosting leading AI firms.

On the competitiveness side, the AI Continent Action Plan sets out an ambitious vision for European AI and identifies five domains that will determine Europe's future position. The InvestAI initiative aims to mobilise \texteuro{}200bn, including \texteuro{}7bn in new public funding, for AI development. The AI Gigafactories initiative plans to establish three to five facilities with at least 100,000 advanced chips each by 2027--8.\footnote{European Commission (2025c) and European Commission (2024a), respectively.} The AI in Science Strategy creates the Resource for AI Science in Europe, inspired by the CERN for AI model, to support scientific applications of AI and improve understanding of AI systems.\footnote{European Commission (2025g).} The Apply AI Strategy focuses on accelerating adoption and contains concrete measures to increase uptake in key sectors.\footnote{European Commission (2025b).} New announcements following the Summit on European Digital Sovereignty in November 2025 indicate that Europe is also betting on frontier systems through a well-funded public-private `Frontier AI Initiative'.\footnote{Elysee (2025).}

Taken together, these initiatives address important gaps in Europe's AI ecosystem and signal growing ambition and scope. However, significant challenges remain once these strategies are calibrated against AGI timelines, both in terms of the scale of potential disruption and the allocation of responsibilities between authorities.

\subsubsection{Existing EU AI strategies are increasingly ambitious, but not yet matched to the pace and scale of global AI competition}

Current EU AI initiatives, notably the AI Continent Action Plan and linked strategies such as Apply AI, InvestAI and AI Factories, are well-scoped and cover the core inputs to AI development. At times, however, they are not fully calibrated to the pace and scale of competition among global AI powers. The buildout of European compute infrastructure is one example. The planned AI Gigafactories represent a substantial step, but they do not close the scale gap between the EU and the US and China.\footnote{We recognise that significant questions remain regarding what the EU's localised compute capacity should be.} The five planned facilities, together providing about 500,000 advanced chips, remain far smaller than comparable US infrastructure.\footnote{European Commission (2024a).} OpenAI alone has said it expects to own well over one million GPUs by the end of 2025, although its facilities are almost entirely privately owned.\footnote{Nasir (2025).}

Recent European strategies push for coordinated action on building and adopting frontier AI models. In the Apply AI strategy, the Commission outlines a `Frontier AI Initiative' that would bring industrial and academic actors together, use AI Factories and Gigafactories, run EU-wide competitions to develop open frontier models and grant winning projects free access to EuroHPC supercomputers. The November 2025 Franco German Digital Sovereignty Summit announced an identically named but distinct `Frontier AI Initiative' that marks a potentially momentous shift in Europe's approach to AI competition. Whereas earlier EU efforts focused mainly on infrastructure and sectoral adoption, this initiative explicitly targets capability competition at the frontier and aims to attract and retain world-class talent able to develop state-of-the-art systems. As of November 2025, no further details on either initiative has been disclosed.

These strategies demonstrate a clear and growing ambition from the Commission, but the implementation path is still uncertain. The \texteuro{}200bn InvestAI target depends heavily on private sector funding, with only about \texteuro{}7bn in new public money committed so far.\footnote{Caroli (2025).} Major delivery questions also surround the Apply AI Frontier AI Initiative. No concrete budget commitments have been published, and the gap between the initiative's planned establishment in the first quarter of 2026 and Gigafactory availability in 2027--8 means recruited teams would initially rely on existing AI Factory infrastructure rather than the frontier-scale compute used by leading frontier AI companies today. A key test of political priority will come with the draft Multiannual Financial Framework for 2028--34, which could signal high-level intent on preparing Europe for a world with AGI.

\subsubsection{Existing EU AI strategies lack the integration required for addressing AGI}

AGI preparedness requires a coherent policy framework that links interventions across several domains: economic competitiveness, technological sovereignty, safety, defence and security, and international governance. It must also cover both EU institutions and Member States. This integration is essential because AGI is likely to cut across traditional policy boundaries in ways that demand coordinated action that no single institution or country can deliver alone. A comprehensive strategy should spell out which values and objectives will sometimes be traded against others, and explain why particular balances are appropriate given the scale of potential disruption. It should also ensure that key institutions work from a shared set of assumptions about AI trajectories and risks.

Current European AI strategies remain spread across several Commission directorates and Member States, with limited coordination. The AI Continent Action Plan distributes responsibilities across multiple initiatives, including ApplyAI,\footnote{European Commission (2025b).} the Data Union Strategy,\footnote{European Commission (2025e).} InvestAI\footnote{European Commission (2025c).} and EuroHPC's AI Factories.\footnote{European Commission (2024a).} This reflects the cross-cutting nature of AI, but also creates interdependencies in which success depends on the coordinated delivery of many sub-strategies, a structure that is likely to slow implementation.

The EU's institutional reliance on consensus, especially in the European Council, adds further friction in moments of geopolitical crisis. For example, Hungary's veto of an \texteuro{}18bn aid package for Ukraine delayed support and forced the EU to improvise alternative financing, showing how unanimity rules in foreign and fiscal policy allow a single Member State to stall collective action.\footnote{Tamma (2022).} Similar dynamics could be damaging in AI competition, where windows for effective policy action are narrow. While AI regulation itself falls under qualified majority voting,\footnote{Qualified majority is the most widely used voting method in the Council. It is used when the Council takes decisions during the ordinary legislative procedure, also known as co-decision. About 80 per cent of all EU legislation is adopted in this way. The Council must vote unanimously on a number of matters which the member states consider to be sensitive (e.g.\ common foreign and security policy, citizenship, EU membership, harmonisation of national legislation on indirect taxation, EU finances, certain provisions in the field of justice and home affairs, and harmonisation of national legislation in the field of social security and social protection).} the fiscal measures needed to accelerate compute buildout or offer tax incentives to attract global talent touch on budget and taxation powers that still require unanimity. This potential internal division also hampers responses to geopolitical coercion. Recent hesitation under US trade pressure has exposed a familiar paralysis.\footnote{Caulcutt \& Gijs (2025).} Even with new tools such as the Anti-Coercion Instrument, divergent economic interests often blunt the political will needed to deploy countermeasures, which limits the EU's collective leverage.\footnote{European Commission (2024).}

\subsubsection{Conclusion: Europe is not adequately prepared for AGI}

Our assessment suggests that Europe faces serious challenges in preparing for a possible transition to AGI across all three criteria in this report. On strategic awareness, Europe trails leading actors, with institutions such as the EU AI Office and INESIA operating with fewer resources and a less focused mandate than the UK's AI Safety Institute. On competitive positioning, Europe trails on most relevant metrics: gaps of about 6 to 12 months in frontier model capabilities, control of only around 5 per cent of global AI compute, attraction of about 6 per cent of global AI venture capital, structurally higher energy prices, and persistent talent retention problems. On leverage, Europe can draw on its role in semiconductor manufacturing equipment and on the weight of the Single Market, yet these advantages are limited and require a more deliberate strategy to translate into sustained geopolitical influence. Finally, existing strategies remain fragmented across institutions and policy domains and often lack the integration and scale that the potential importance of AGI would require.

\section{Recommendation: the European Commission President should commission an AGI Preparedness Report}
\label{sec:recommendation}

Our evaluation of the available evidence suggests that AGI may arrive soon, plausibly between 2030 and 2040, or even earlier. We identify pathways via which the emergence of AGI could reshape the global distribution of power, with major implications for economic growth, military capabilities and international stability. We find that Europe is not adequately prepared, lacking the strategic awareness, competitive positioning and policy strategies to navigate a transition to AGI.

If the EU is to capture AGI's economic benefits, become more resilient to external coercion and ensure safe and responsible development and deployment, it needs a plan. To begin meeting this challenge, we offer the following recommendations:

\begin{enumerate}
  \item The President of the European Commission should commission an `AGI Preparedness Report' addressing three questions:
  \begin{enumerate}
    \item How can Europe capture economic benefits from AGI while remaining sovereign?
    \item How can Europe prepare its institutions and societies for rapid change from AGI?
    \item How can Europe strengthen international stability, security, and shared prosperity in a world shaped by AGI?
  \end{enumerate}
  \item The Report should be spearheaded by a lead author with both strong technical credibility and the highest political standing. The lead author should be afforded:
  \begin{enumerate}
    \item The intellectual freedom to write an independent analysis.
    \item Resources to assemble a small, agile, team of world-class experts to conduct technically rigorous research, and commission analyses from external experts.
    \item Senior political support to move through processes fast, including in terms of input from EU institutions, Member State governments and external actors.
  \end{enumerate}
  \item The Report should be technically rigorous and based on the best available evidence, while also taking an anticipatory stance and being explicit about uncertainty.
  \item The Report should be delivered quickly, ideally within six months, to reflect that AGI may be near and to maximise the available time for implementation.
  \item The Report should be presented to the European Council at a dedicated event, to enable a parallel diplomatic process and high-level endorsements.
\end{enumerate}

Below we explain why Europe needs a rigorous AGI preparedness plan before adopting a series of far-reaching policies. We then show why the Draghi Report on EU Competitiveness offers a particularly fitting blueprint for an AGI Preparedness Report but also highlight areas where the latter would have to go beyond Draghi's approach. Finally, we set out the key strategic questions that the Report should address.

\subsection{Why a report instead of concrete action recommendations?}

It may appear overly slow and bureaucratic for the EU to first commission an elaborate report before implementing concrete policies on AGI preparedness. Some might argue that the Draghi report itself spurred little practical action---according to the Draghi Observatory and Implementation Index, only 43 (11.2 per cent) of the Report's 383 recommendations have been fully implemented a year on.\footnote{Bulski (2025).} However, we argue that a report is the correct course of action, for three reasons:

\begin{itemize}
  \item \textbf{Firstly, the stakes are so high that policy responses of the right magnitude will often be extremely consequential and getting them wrong would be costly financially and (geo)politically.} For example, spending hundreds of billions of euros on datacentres and energy infrastructure might seem essential today, but could prove wasteful if technical and economic paradigms at the AI frontier shift. Moreover, the EU might only get a few `shots' to get it right. If we assume plausible AGI timelines between 2030 and 2040, or even earlier, then decisive action is necessary soon and there might be little time to course-correct.

  \item \textbf{Secondly, AGI could cut across policy domains more deeply than any previous technology.} Everything from economic policy, to security policy, and from social policy to foreign policy may be affected and there will be both important synergies and significant trade-offs between these domains. AGI preparedness therefore requires a coherent policy portfolio that reflects these interdependencies and is anchored in a shared set of assumptions and strategic priorities. This calls for structured involvement of EU institutions and Member States, industry, labour, academia, organisations such as NATO, security and intelligence services, sectoral regulators, and civil society. The need for integration makes an EU led sense making effort more appropriate than a purely national initiative.

  \item \textbf{Thirdly, the technical details will matter.} For example, the shift from scaling pre-training compute to scaling inference compute could alter the AI governance landscape and with it the strategic logic of a European policy response.\footnote{Ord (2025).} Future technical developments may have similar effects. Europe's options for AGI preparedness must therefore be assessed with the utmost scientific rigour and with input from AI experts, including eminent scientists and engineers from leading frontier AI companies.
\end{itemize}

Taken together, an in-depth, comprehensive analysis, produced through a rigorous and authoritative process is required in order to become better prepared for AGI. The initial focus should be on clarifying the big picture, identifying the most important strategic questions, mapping the option space for addressing them, and outlining high-level directions and trade-offs. Once the priorities are clear, Europe can move towards policy implementation.

\begin{tcolorbox}[colback=gray!5,colframe=gray!50,title=How an AGI Preparedness Report would differ from related reports]
\textbf{Compared with the EU Strategic Foresight Report,}\addtocounter{footnote}{1}\textsuperscript{\thefootnote} an AGI Preparedness Report would be narrower and more action-guiding. While the EU Strategic Foresight Report examines broad trends across multiple domains with a 10--20 year horizon, an AGI Preparedness Report would focus specifically on the challenges posed by the potential near-term emergence of AGI. It would be operationally urgent rather than exploratory, delivered within six months rather than as part of an annual cycle, and structured around immediate policy choices rather than long-term scenario planning. Most importantly, it would identify trade-offs and name concrete owners and timelines for action, whereas the EU Strategic Foresight Report typically preserves more flexibility.

\textbf{Compared with the International AI Safety Report,}\addtocounter{footnote}{1}\textsuperscript{\thefootnote} an AGI Preparedness Report would be more anchored in EU policy and geopolitical choices. While the International AI Safety Report provides valuable technical consensus on frontier AI capabilities, risks and safety measures, drawing on expertise from 30 countries and international organisations, it does not focus on any particular country or region. An AGI Preparedness Report would focus on Europe's specific position in AGI geopolitics, such as its 6--12 month lag in frontier capabilities, its dependencies on US cloud infrastructure and chips, and its limited strategic awareness capacity. It could identify leverage points such as ASML's chokepoint role, Single Market access, and regulatory expertise, then propose strategies for converting these into assured access to AGI capabilities and influence over global development. It would also need to navigate EU and Member State competency divisions, clarify what requires Brussels level coordination and what demands national leadership, and propose reforms suited to European decision making structures.

\textbf{Compared with the EU Strategic Compass for Security and Defence,}\addtocounter{footnote}{1}\textsuperscript{\thefootnote} an AGI Preparedness Report would span a broader set of challenges beyond traditional security and defence. While the Strategic Compass provides essential guidance for crisis management, military capabilities, defence industrial policy and security partnerships, organised around Act, Secure, Invest, and Partner, it does not treat AGI as a distinct strategic challenge. An AGI Preparedness Report would integrate the security and defence dimensions addressed by the Compass with economic competitiveness, technological sovereignty and societal resilience.
\end{tcolorbox}
\addtocounter{footnote}{-3}
\stepcounter{footnote}\footnotetext{European Commission (2025h).}
\stepcounter{footnote}\footnotetext{Bengio et al.\ (2025).}
\stepcounter{footnote}\footnotetext{European External Action Service (2024).}

\subsection{Learning from the Draghi Report on EU Competitiveness}

The Draghi Report\footnote{Draghi (2024).} showed how high-level political backing combined with rigorous technical analysis, can shift policy debates in areas where political attention lags behind geopolitical reality. Its impact rested on a clear mandate from the Commission President, an expert process led by a highly authoritative figure that produced actionable proposals, and a public launch that made the status quo harder to defend.

A report cannot override Member State sovereignty or Commission competences, but it can still reshape the terms of debate, create benchmarks for accountability, and raise the political cost of fragmented responses to a problem requiring coordination. Although many of Draghi's recommendations remain unimplemented and some may already be dated, the report established a framework that now structures how competitiveness is discussed and measured across European institutions. An AGI Preparedness Report should aim for similar effects by creating common vocabulary, establishing hard to dispute facts, and giving political cover to leaders who recognise the urgency and need authoritative backing for exceptional measures.

\subsubsection{Where an AGI Preparedness Report should take inspiration from Draghi}

The Draghi Report's impact was the result of deliberate choices: strategic focus, proportionate ambition, intellectual independence, operational concreteness, EU and Member State integration and execution by a diverse, agile team with high-level political backing. An AGI Preparedness Report should draw on these lessons while adapting them to AGI's distinctive challenges.

\paragraph{Strategic focus.} The competitiveness report kept a narrow scope centred on revitalising European competitiveness. This enabled deeper analysis of root causes and created space for politically difficult proposals that a broader approach might have avoided. The AGI Preparedness Report should similarly focus on AGI's implications for Europe's core interests, including economic prosperity, sovereignty, security, and international stability, rather than attempting comprehensive AI policy coverage. It should explicitly address scenarios in which AGI emerges within the next decade, recognising that preparation for faster timelines builds capacity that remains valuable even if progress slows, while the reverse leaves Europe dangerously exposed.

\paragraph{Proportionate ambition.} The competitiveness report argued that incremental adjustments would not be sufficient and proposed politically unpalatable measures, such as massive investment increases and coordination that required Member States to cede autonomy. The AGI Preparedness Report must also match its recommendations to the scale of the AGI challenge. It should assess options that may seem disproportionate under normal conditions, such as multi hundred billion euro investments and faster EU decision processes, while naming the tradeoffs clearly. It should also reflect a basic reality that Europe cannot simultaneously maximise innovation, minimise all risks, preserve complete privacy, maintain full sovereignty, and avoid significant investment.

\paragraph{Operational concreteness.} The competitiveness report offered actionable recommendations with clear owners and timelines. The AGI Preparedness Report should do the same by specifying concrete actions with clear owners, realistic timelines, and identified resources. Where action is needed but details require further work, it should call for dedicated working groups and define specific deliverables. The aim is high level clarity on what must happen, who is responsible, and what success looks like.

\paragraph{Intellectual independence and political mandate.} The competitiveness report carried the authority of a former Prime Minister and ECB President and had direct backing from the Commission President. This gave Draghi the convening power to engage heads of state and industry leaders and made the findings harder to dismiss, while preserving room for politically costly recommendations. The AGI Preparedness Report needs the same balance. The lead author should be shielded from political and industry pressure, authorised to reach uncomfortable conclusions, and report directly to the Commission President.

\paragraph{A diverse, agile team.} The competitiveness report drew on a dedicated Commission team of 11 expert staff with deep institutional knowledge of EU economics, policy, and governance supported by extensive consultation of prominent external academics, including Nobel laureates. The AGI Preparedness Report must ensure it includes experts who understand the specifics of AI and AGI, including specialists currently working at frontier AI companies. The team should therefore include technical AI researchers familiar with the capabilities frontier, economists focused on the economic impacts of advanced AI, geopolitical and security experts who can assess technological capabilities and their consequences for international competition, and legal and governance specialists who understand the regulatory challenges posed by transformative AI systems.

Assembling such a team would likely require departing from standard EU hiring practices. The lead author should be authorised to offer compensation competitive with private sector opportunities, potentially several multiples of normal EU salary scales.

\paragraph{Integration across EU and Member States.} The competitiveness report recognises that many levers sit with Member States but require coordination at European scale. The AGI Preparedness Report must go further. Many AGI relevant levers sit in domains where Member State sovereignty remains central, including defence and security, intelligence cooperation, and critical infrastructure protection. The report should therefore specify which actions require Brussels level coordination, such as unified positions in international AI negotiations, semiconductor supply chain security, and standards for shared compute, and which require national leadership, such as military AI applications, domestic security systems, and classified capability development. It should also define mechanisms that ensure both levels act in concert rather than in ways that conflict with each other.

This integration task is politically sharper than the one Draghi faced. Member States will differ in priorities, from industrial policy to fundamental research, from closer US partnership to stronger European autonomy, and from willingness to fund large public investments to fiscal restraint. An AGI Preparedness Report should acknowledge these differences while building agreement on a core agenda that allows coordinated action, even in areas traditionally considered too sensitive for deep Brussels involvement.

\subsection{Core questions an AGI Preparedness Report should address}

This section does not prescribe policy recommendations---these should emerge from an expert-led process. It sets out the core strategic questions the AGI Preparedness Report must address, grouped into three challenges: capturing economic benefits while remaining sovereign; preparing European society and institutions for rapid change; and strengthening international stability, security, as well as shared prosperity.

\subsubsection{1. How can Europe capture economic benefits from AGI while remaining sovereign?}

Europe faces consequential choices about capturing AGI's economic value without dependencies that could enable external coercion. This tradeoff between access and sovereignty will shape decisions from infrastructure investments to industrial strategy to supply chain positioning. Reliance on US and Chinese systems increases coercion risks, while full autarky would require massive long-term and high-risk investment, potentially locking Europe into sustained technological disadvantage.

\paragraph{How should Europe define and operationalise AI sovereignty?}
Different applications require different levels of assured access and control. Military, intelligence, and critical infrastructure likely require full European control, while less sensitive domains may accept controlled dependence if continuity and security are credibly assured. The central choice is when to procure AI capabilities from foreign providers and when to build them domestically, since procurement can create strategic dependencies and domestic development can be slow and costly.

\begin{itemize}
  \item Which application areas require full European sovereign control? Where is controlled dependence acceptable and when is standard commercial procurement sufficient? What criteria should define these tiers?\footnote{For example, are current rubrics such as the Sovereignty Effectiveness Assurance Levels (SEAL) classification found in the EU Cloud Sovereignty Framework sufficiently robust?; European Commission (2025l).}
  \item Which technical, legal, operational and political assurances are required for each tier? How can these be designed to guarantee access to frontier capabilities and compute amid geopolitical turbulence?
  \item How should procurement be structured? For example, how much reliance on US systems is acceptable given the demonstrated use of technology as a geopolitical lever? What limits should govern engagement with Chinese providers? Under which conditions is it acceptable to rely on open-weight models?
\end{itemize}

\paragraph{What infrastructure does Europe need for AI sovereignty, and how can it be built fast enough?}
Europe's AI sovereignty objectives imply concrete requirements for AI infrastructure. These vary by workload: frontier training needs hyperscale, power-dense clusters with tight interconnect; high-security uses require air-gapped, physically isolated facilities, and low-latency inference calls for distributed capacity across regions. Sound choices will depend on Europe's energy availability and security and on the strength of its tech base especially AI and cloud services.

\begin{itemize}
  \item How much datacentre and energy infrastructure does Europe need to meet its sovereignty objectives? What characteristics should this infrastructure have across training versus inference, security requirements, and geographical centralisation? What mix of public, private and foreign ownership and financing fits different sovereignty objectives?
  \item How can Europe expand its datacentre infrastructure and grid capacity expand at a speed that meets its sovereignty objectives? Where should datacentres be built to maintain access to reliable, cheap and non-intermittent power?
  \item Is it necessary and feasible for Europe to build frontier AI models? How close can and should European models be to the frontier, and how can they be developed safely? What effort, cost, and timeline would be required at a given level of ambition?
  \item If Europe decides to build frontier models, how should it do this? Should Europe support national AI champions, pursue bilateral collaborations, organise multilateral middle-power efforts with countries such as the UK, Canada, Japan and South Korea, or seek partnership in U.S.-led frontier AI projects?\footnote{For example, joining a potential multilateral public-private partnership between the US government and a frontier AI company.} What balance between public and private sector engagement is ideal?
\end{itemize}

\paragraph{How can Europe secure access to key input factors?}
Even with some sovereign capabilities, Europe remains vulnerable when critical dependencies create opportunities for external coercion. The most acute near-term risk may be US restrictions on access to advanced semiconductor, especially NVIDIA's leading AI chips, since a cutoff would quickly constrain European AI development.

\begin{itemize}
  \item How should Europe hedge against potential US chip restrictions while maintaining productive relationships with US technology partners? Should Europe use control over key semiconductor supply chain assets such as ASML as a lever to secure access to foreign chips? Is this technically, legally and politically feasible? Under what conditions would this be constructive rather than escalatory?
  \item How can Europe strengthen its position across the AI supply chain? Should it indigenise more of the existing supply chain and, if so, how? Should it strategically invest in emerging AI stack technologies or adjacent non-AI technologies that could provide future leverage? Which European manufacturing and industrial capabilities might become more valuable as AI enables physical automation, and how can Europe reinforce them? How can Europe prepare its institutions and societies for rapid change from AGI?
\end{itemize}

\subsubsection{2. How can Europe prepare its institutions and societies for rapid change from AGI?}

Europe must build resilience against the disruptions AGI may cause, from labour market upheavals to novel security threats and strains on governance and free elections. Given the pace of frontier AI development and uncertainty about the potential implications of AGI, European institutions must be able to adapt quickly while maintaining democratic legitimacy.

\paragraph{How can European governments and institutions maintain strategic awareness on frontier AI trends?}
AGI preparedness requires that Member State governments and EU institutions stay informed about recent technical and geopolitical trends in frontier AI, and how these trends might affect policy objectives. This strategic awareness depends on strong technical capacity in government, robust policy expertise, and institutional channels that translate findings into timely guidance to policymakers.

\begin{itemize}
  \item Which new and upgraded EU and Member State institutions are needed to ensure adequate strategic awareness? For example, should the EU AI Office build a strategic awareness function comparable to that at the UK AI Security Institute, and should more Member States establish their own AISIs and strategic awareness teams? What mandates would ensure effective coordination without duplication or interference?
  \item What institutional models would help governments attract and retain the required expertise despite private sector competition? For example, what flexibility on pay, hiring procedures and autonomy would be needed to persuade a leading AI scientist in Silicon Valley to join a European strategic awareness unit?
\end{itemize}

\paragraph{How can European governments and institutions retain agency amidst fast technological change?}
If AGI arrives in the near future, Europe will need faster decision making. This could include responding to large scale AI accidents or loss of control incidents, accelerating infrastructure investments, and resisting foreign technological coercion. Standard EU policy processes often move on multi-year timescales poorly matched to technology that can shift in months. Without institutional adaptation, even well-designed policies may fail in implementation or arrive too late to matter.

\begin{itemize}
  \item How can the EU and Member States ensure that AGI preparedness decisions can be made at the right speed? Which institutional mechanisms and reforms can deliver faster action across policy domains while preserving democratic accountability? What should EU and Member State coordination look like in this context?
  \item When should accelerated decision mechanisms be activated? For example, should triggers be tied to capability milestones or impact indicators such as models showing sustained autonomy on month-long tasks, evidence of widespread labour market disruption or credible demonstrations of catastrophic misuse potential?
  \item How should Member State governments and EU institutions use AGI to improve their own decision making and remain competitive with other AGI enabled actors?\footnote{For example, private sector companies using AGI for automated lawsuits, governments using AGI for negotiations.} How can democratic legitimacy be protected when governance is increasingly supported by AGI?
  \item What contingency plan does Europe need if AGI arrives very quickly?\footnote{Predd (2025).} If AGI does not arrive quickly, how can Europe avoid wasting resources on premature preparation?
\end{itemize}

\paragraph{How can Europe defend its institutions and societies against AGI-enabled threats?}
As AI capabilities advance, Europe faces rising exposure to new forms of attack. Current systems already enable more sophisticated cyberattacks and have shown potential for misuse in biological design. The same methods that accelerate protein folding, drug discovery and genetic engineering could also lower the barriers that once kept pathogen design largely confined to trusted labs. Europe must ensure that regulators, the judiciary, security authorities, industry, and civil society are prepared to detect, deter, and respond to these threats.

\begin{itemize}
  \item How can Europe ensure that the EU AI Act and the Code of Practice for general purpose AI remain credible and enforceable as capabilities scale and competitive pressures intensify? How can the Commission ensure the EU AI Office has the resources and flexibility needed for effective enforcement?
  \item How can Europe promote the Code of Practice internationally to spread state-of-the-art risk assessment, evaluation, and mitigation beyond Europe?
  \item How can Europe ensure preparedness across the full range of EU and Member State institutions responsible for defending against AGI enabled threats, including militaries, intelligence agencies, cybersecurity units, pandemic preparedness bodies, and law enforcement? What expertise, technologies and protocols would these institutions need to function effectively in an AGI crisis?
  \item What resilience measures should Europe introduce to prepare civil society for AGI?\footnote{For example, AI-literacy programs aimed at reducing susceptibility to AGI-based foreign influence operations.}
  \item Which defensive technologies relating to AGI safety, security and transparency should Europe foster, and what kinds of research bodies are best placed to develop them?\footnote{For example, AI defences against cyberattacks on critical infrastructure, advanced methods for pandemic pathogen detection, and verification mechanisms that allow enforcement of international treaties.}\textsuperscript{,}\footnote{For example, breakthrough innovation agencies such as SPRIND and ARPA, Focused Research Organisations.}
\end{itemize}

\paragraph{How can Europe protect its social and economic model against AGI-based disruption?}
AGI will likely boost economic growth, but it may also displace workers and erode traditional tax bases. Europe may need new taxation and distribution mechanisms to sustain the welfare state and other core public functions in a highly automated economy.

\begin{itemize}
  \item How can Europe capture productivity gains from AGI-enabled automation while providing strong support for displaced workers? How should programmes and incentives for upskilling, retention, and retraining evolve as automation deepens?
  \item How can governments respond if labour income tax bases erode? How can they ensure AGI enabled growth translates into public revenue rather than concentrating private wealth?
  \item How can Europe preserve democratic checks and balances if citizens' labour becomes less central to economic value?
\end{itemize}

\subsubsection{3. How can Europe strengthen international stability, security and shared prosperity in a world shaped by AGI?}

Europe can help shape a stable, rules based international order for responsible AGI development while safeguarding European interests amid intensifying geopolitical competition. Europe's position outside the US-China rivalry creates both opportunities and constraints that require careful navigation.

\paragraph{How can Europe influence global AGI governance if it does not host frontier AI companies?}
The US and China have a direct path to shape AGI governance because they host the leading AI firms and much of the world's computing capacity. Unless Europe builds frontier AI champions through industry or public private partnerships, its influence will be more indirect and will rely more on indirect means such as market access.

\begin{itemize}
  \item Which levers beyond market access can Europe use to shape global AGI governance? For example, can trade agreements, procurement rules, and control over access to compute, energy and key semiconductor supply chain assets function as bargaining chips? Are these options technically, legally and politically feasible, and under what conditions should each be used?
  \item How can Europe maximise its geopolitical leverage over the long term? Should it invest strategically in novel technologies and applications that make Europe indispensable in a world with AGI?
  \item Which diplomatic alliances should Europe build to increase its influence on global AI governance? For example:
  \begin{itemize}
    \item In which areas should Europe collaborate with the US, and when might strategic ambiguity toward China be useful?
    \item How should Europe collaborate with AI bridge powers\footnote{Abecassis et al.\ (2025). `AI bridge powers' refers to countries which are technologically advanced but share Europe's predicament of possessing significant capabilities while lacking autonomous access to frontier development.} such as the UK, Canada, Japan, South Korea, and India, including on shared compute infrastructure, pooled supply chain bargaining power, joint research programmes and coordinated diplomatic positions? To what extent should the EU engage with Middle Eastern partners?
  \end{itemize}
  \item How can Europe shape global AGI governance by providing public goods such as verification technologies? For example, could Europe develop hardware-based verification tools that enable credible international agreements between the US and China despite mutual distrust?
\end{itemize}

\paragraph{How can Europe help create a stable and broadly beneficial global AGI governance regime?}
Beyond managing immediate competitive pressures, Europe must articulate a long-term vision for global AGI governance. This vision should serve as an aspirational goal and a guiding frame for near term diplomatic initiatives.

\begin{itemize}
  \item What is Europe's vision for a stable international order in a world with AGI? For example, should it favour a market led model, advocate a global pause beyond defined risk thresholds, support a United States bid for unilateral advantage, work toward a single internationalised AGI project, pursue a pluralistic system with a globally coordinated regulator, or another approach?
  \item What concrete steps can Europe take to advance this vision? Which governance norms and technical standards should it champion globally, and which partnerships, alliances, and institutions are needed to uphold them? Could Europe help facilitate a productive US-China dialogue on AI safety, or offer technical cooperation that both powers might accept through neutral European institutions?
  \item What is Europe's vision for how the economic benefits from AGI should be distributed globally? How can fundamental human rights be protected in a world with AGI?
\end{itemize}

\section{Conclusion}
\label{sec:conclusion}

This report has examined the evolving geopolitics of AGI through a European lens, tracing the trajectory from technological possibility to geopolitical imperative. It concluded that Europe must urgently rethink its strategic approach to AGI.

The three preceding chapters have laid out the strategic landscape:

\begin{itemize}
  \item Despite the jagged shape of the current AI frontier, trends in the underlying input factors indicate that further advances in capabilities are possible. AI systems that can automate most economically valuable cognitive work---our working definition of AGI---may arrive soon, plausibly between 2030 and 2040, or even earlier. While this timeline is uncertain, the likelihood is sufficient to warrant significant attention by policy makers.

  \item States that successfully leverage AGI may achieve unprecedented economic growth through comprehensive labour automation and accelerated R\&D. This could enable them to outgrow and progressively marginalise non-AGI powers. In the military domain, AGI could open novel warfare paradigms that render existing defence frameworks obsolete. Moreover, AGI may empower non-state actors to do significant harm via mechanisms such as human uplift for weapons of mass destruction, misaligned rogue AI, or extreme concentration of power.

  \item Europe lacks adequate preparedness for a potential transition to AGI. The continent trails US and Chinese frontier models by 6--12 months, controls only about 5 per cent of global AI compute, attracts just 6 per cent of global AI venture funding, and faces persistent challenges in talent retention. It maintains leverage through ASML's lithography monopoly and its substantial market power, but these positions are constrained by structural dependencies and risks of retaliation. Its AI strategies remain fragmented across domains and institutions.
\end{itemize}

Together, these factors place Europe in a difficult position. AGI could reshape economic prosperity, military power and political order within the next decade, but Europe is currently not ready to navigate this shift. The window for action may be narrow. Decisions made over the next few years will determine Europe's future geopolitical position and its ability to guide AGI development toward safer and more responsible outcomes.

Still, Europe retains significant strength. The EU's economic scale, regulatory capacity, research base and commitment to multilateral cooperation provide a strong foundation. By building on this strength, Europe can assume a leading role in an AGI transition. This work begins with a plan:

\begin{enumerate}
  \item The President of the European Commission should commission an `AGI Preparedness Report' addressing three questions:
  \begin{enumerate}
    \item How can Europe capture economic benefits from AGI while remaining sovereign?
    \item How can Europe prepare its institutions and societies for rapid change from AGI?
    \item How can Europe strengthen international stability, security, and shared prosperity in a world shaped by AGI?
  \end{enumerate}
  \item The Report should be spearheaded by a lead author with both strong technical credibility and the highest political standing. The lead author should be afforded:
  \begin{enumerate}
    \item The intellectual freedom to write an independent analysis.
    \item Resources to assemble a small, agile, team of world-class experts to conduct technically rigorous research, and commission analyses from external experts.
    \item Senior political support to move through processes fast, including in terms of input from EU institutions, Member State governments and external actors.
  \end{enumerate}
  \item The Report should be technically rigorous and based on the best available evidence, while also taking an anticipatory stance and being explicit about uncertainty.
  \item The Report should be delivered quickly, ideally within six months, to reflect that AGI may be near and to maximise the available time for implementation.
  \item The Report should be presented to the European Council at a dedicated event, to enable a parallel diplomatic process and high-level endorsements.
\end{enumerate}

\appendix
\section{Glossary}
\label{app:glossary}

\begin{description}[style=nextline, leftmargin=0pt, labelindent=0pt]
\item[Agentic AI] AI systems that can plan, act, and adapt to pursue goals across multi-step tasks with limited human oversight.
\item[AI Continent Action Plan] A 2025 European Commission agenda to scale AI adoption, skills and infrastructure while supporting AI Act implementation.
\item[AI Datacentres] Facilities with dense compute, storage, and networking optimised for large-scale AI training and inference.
\item[AI Diffusion] The spread of AI adoption and use across sectors, firms and public institutions.
\item[AI Gigafactory] Large EU-backed AI compute sites intended to provide frontier-scale training and deployment capacity in Europe.
\item[AI in Science Strategy (RAISE)] A 2025 European Commission strategy to accelerate AI for scientific discovery through pooling data, compute and talent via a `Resource for AI Science in Europe' (RAISE).
\item[AI Misuse] The intentional use of AI systems to cause harm, enable wrongdoing or violate laws and norms.
\item[AI Office (EU AI Office)] The European Commission body coordinating AI Act implementation and oversight, including for general-purpose AI models, alongside national authorities.
\item[AI Triad] The three core drivers of recent AI progress: data, compute and algorithms.
\item[Algorithmic Efficiency] Improvements that increase AI performance per unit of compute and data.
\item[Alignment] Making AI systems reliably behave in line with the designer's intentions and values.
\item[API (Application Programming Interface)] A software interface for accessing model capabilities programmatically without direct access to model weights.
\item[Apply AI Strategy] A 2025 EU sectoral strategy to accelerate AI adoption across key industries and the public sector, with strong emphasis on SMEs.
\item[Brussels Effect] The tendency for EU rules to become de facto global standards as firms align worldwide practices with EU compliance.
\item[Chain-of-Thought] A prompting/training approach that elicits intermediate reasoning text to improve problem-solving and explain outputs.
\item[Chokepoint Technologies] Hard-to-replace technologies or components that confer strategic leverage due to concentrated supply or know-how.
\item[Closed-weight Models] Models whose weights are not publicly released and are typically accessed only through products or APIs.
\item[Code of Practice (for GPAI)] A voluntary framework helping GPAI providers demonstrate alignment with relevant AI Act obligations.
\item[Compute / Computing Power] The processing capacity required to train, fine-tune and run AI models.
\item[Context Window] The maximum amount of input information a model can attend to at once when generating output.
\item[Data Wall] The hypothesised point where progress slows because high-quality training data becomes scarce or costly to obtain.
\item[Dual-use AI] AI capabilities that can deliver major benefits but can also be repurposed for harmful or illicit ends.
\item[Effective Compute] The equivalent increase in computational scale that would be needed to match a given model performance absent algorithmic innovation.
\item[Embodied AI] AI integrated into physical agents, such as robots, vehicles, devices, that perceive and act in the real world.
\item[EU AI Act] The EU's risk-based AI law adopted in 2024, setting obligations across prohibited, high-risk, and general-purpose AI uses.
\item[Fine-tuning] Additional training that adapts a pre-trained model to specific tasks, domains or behaviours.
\item[FLOP (Floating Point Operations)] A measure of computational work; exaFLOP = $10^{18}$ FLOP, and frontier training is often discussed in ${\sim}10^{25}$--$10^{26}$ FLOP ranges.
\item[Frontier AI Models] The most capable models at the cutting edge of performance and potential risk.
\item[General-purpose Models (GPAI)] Broadly capable models designed to perform many tasks across domains rather than a single narrow use.
\item[GPQA Diamond] A benchmark of difficult graduate-level science questions used to test advanced model reasoning.
\item[GPU (Graphic Processing Unit)] Highly parallel chips widely used for AI training and inference.
\item[HBM (High-Bandwidth Memory)] Advanced stacked memory critical for feeding data fast enough to modern AI accelerators.
\item[Hyperscaler] A cloud provider operating massive data centres that can scale AI training and deployment globally.
\item[Inference] Running a trained model to generate outputs from new inputs.
\item[Inference-time Scaling Laws] The idea that more test-time compute, such as longer deliberation, can improve model performance.
\item[Interpretability / Mechanistic Interpretability] Methods for understanding and auditing how models represent knowledge and produce decisions.
\item[Large Language Model (LLM)] A model trained on large text corpora to generate, transform and reason over language.
\item[Latency] The time delay between providing an input and receiving a model's output.
\item[Lithography (EUV / DUV)] Semiconductor manufacturing techniques used to pattern advanced chips at increasingly small feature sizes.
\item[Loss of Control (Active / Passive)] Risks that AI systems pursue unintended goals (active) or become too complex/fast to manage safely (passive).
\item[MAD (Mutually Assured Destruction)] A nuclear-era deterrence concept where full-scale attack ensures catastrophic retaliation.
\item[MAIM (Mutually Assured AI Malfunction)] A proposed AI-era deterrence concept where states threaten mutual disruption of dangerous AI projects.
\item[MMLU] A broad benchmark spanning many subjects used to test general knowledge and reasoning.
\item[Model / Capability Evaluations] Structured tests that measure what a model can do and the risks it may pose in real use.
\item[Model Weights] The learned numerical parameters that store a model's capabilities and knowledge.
\item[Multimodal Models] Models that can process and generate across multiple modalities, such as text, images, audio, or video.
\item[Neural Networks] Machine-learning systems composed of layered units that learn patterns by adjusting weights via training.
\item[Open-weight / open-source models] Models with publicly downloadable weights; `open-source' further implies permissive code/licensing and sometimes broader transparency.
\item[Post-training] Training after pre-training to improve instruction-following, safety or specialised skills, often via fine-tuning and reinforcement learning.
\item[Pre-training] The large-scale initial training phase where models learn general patterns from massive datasets.
\item[Reasoning Models] Models optimised to solve complex problems through longer, structured internal or textual reasoning.
\item[Reinforcement Learning (RL)] A training approach where models learn behaviours through rewards, penalties and iterative feedback.
\item[Reward Hacking] When an AI finds unintended shortcuts to maximise a reward signal without achieving the real objective.
\item[Scaling Laws] Empirical relationships showing how performance tends to improve with more data, compute and model capacity.
\item[Scalable Oversight] Approaches that maintain effective monitoring and control as AI systems become more capable.
\item[Sleeper Agents] Systems that behave benignly in testing but may activate harmful goals under specific triggers or contexts.
\item[Special Compute Zones] Proposed designated regions with streamlined rules, power planning and investment to accelerate large AI cluster deployment.
\item[Synthetic Data] Artificially generated training data used to augment or replace scarce real-world data.
\item[Task Horizon] The length and complexity of tasks an AI can autonomously plan and complete.
\item[Token] A basic unit of text that models process during training and inference.
\item[Training] The process of optimising model weights on data to improve performance on target objectives.
\item[Transformer Architecture] A 2017 neural architecture based on attention mechanisms that underpins most modern LLMs.
\item[US AI Action Plan] A 2025 White House plan centred on accelerating innovation, building AI infrastructure, and leading on international AI diplomacy and security.
\item[Verifier Systems] Models or tools that check, score, or validate candidate outputs to improve reliability and guide learning.
\item[World Foundation Models] Advanced models aimed at learning rich, general representations of the physical and digital world for prediction and planning.
\end{description}

\section*{Acknowledgments}

We sincerely thank the following people for their input and/or written feedback on this report: Anton Leicht, Anton Shenk, Audrey Tang, Chris Meserole, Erich Grunewald, Henry Papadatos, Jonathan Schmidt, Jorge Teixeira, Luka Ignac, Markus Anderljung, Michael Aird, Oscar Delaney, Patricia Paskov, Philip Fox, Rafael Andersson Lipcsey, Saad Siddiqui, Seb Krier, Shyam Krishna, Siddhi Pal, Soeren Mindermann, Stefan Leijnen, Tom Ough. We are particularly grateful to Henri van Soest and James Black for their detailed review and to Jessica Plumridge, Ben Plumridge, Rowan Emslie, Georgina Melia, and Lauren Crawford for copy-editing and design.

These individuals do not necessarily agree with the views in the paper, and all mistakes remain our own. This paper is a collaborative effort. While the authors are aligned on the paper's core argument, individual authors may differ on some specific claims or interpretations.

\paragraph{Funding.} This research was independently initiated and conducted jointly by the Center on AI, Security, and Technology and the Centre for Future Generations using income from operations and gifts and grants from philanthropic supporters. A complete list of RAND's donors and funders is available at \url{https://www.rand.org/CAST}. RAND clients, donors and grantors have no influence over research findings or recommendations. The Centre for Future Generations (CFG) is a nonprofit organisation supported by philanthropic foundations, public institutions, and individual donors. CFG does not accept funding that could compromise its independence, including primary support from governments, corporations, or political parties.

{\sloppy

}

\section*{How to Cite This Publication}

\begin{quote}
Negele, M., Juijn, D., Shamir, A., Jank\r{u}, D., \"{O}zcan, B., Soder, L., Velasco, L., Reddel, M., Bakker, M., Pacchiardi, L., \& Andriushchenko, M. (2025). \textit{Europe and the Geopolitics of AGI: The Need for a Preparedness Plan}. RAND Corporation and Centre for Future Generations. RR-A4636-1.
\end{quote}

\noindent BibTeX:
\begin{verbatim}
@techreport{negele2025geoagi,
  author      = {Negele, Maximilian and Juijn, Daan and
                 Shamir, Afek and Jank{\'u}, David and
                 {\"O}zcan, Beng{\"u}su and Soder, Lisa and
                 Velasco, Lucia and Reddel, Max and
                 Bakker, Michiel and Pacchiardi, Lorenzo and
                 Andriushchenko, Maksym},
  title       = {Europe and the Geopolitics of AGI:
                 The Need for a Preparedness Plan},
  institution = {RAND Corporation and Centre for
                 Future Generations},
  year        = {2025},
  number      = {RR-A4636-1},
  url         = {https://www.rand.org/t/RRA4636-1}
}
\end{verbatim}

\end{document}